\documentclass[pss]{wiley2sp} 
\usepackage{amsmath}
\usepackage{bm}             

\graphicspath{{figures/}}

\begin{document}

\title{Compositional Studies of Metals with Complex Order by means of the Optical Floating-Zone Technique}

\author{%
  Andreas Bauer\textsuperscript{\Ast,\textsf{\bfseries 1}},
  Georg Benka\textsuperscript{\textsf{\bfseries 1}},
  Andreas Neubauer\textsuperscript{\textsf{\bfseries 1}},
  Alexander Regnat\textsuperscript{\textsf{\bfseries 1}},
  Alexander Engelhardt\textsuperscript{\textsf{\bfseries 1}},
  Christoph Resch\textsuperscript{\textsf{\bfseries 1}},
  Sabine Wurmehl\textsuperscript{\textsf{\bfseries 2}},
  Christian G. F. Blum\textsuperscript{\textsf{\bfseries 2}},
  Tim Adams\textsuperscript{\textsf{\bfseries 1}},
  Alfonso Chacon\textsuperscript{\textsf{\bfseries 1}},
  Rainer Jungwirth\textsuperscript{\textsf{\bfseries 3}},
  Robert Georgii\textsuperscript{\textsf{\bfseries 1,3}},
  Anatoliy Senyshyn\textsuperscript{\textsf{\bfseries 3}},
  Bj\"{o}rn Pedersen\textsuperscript{\textsf{\bfseries 3}},
  Martin Meven\textsuperscript{\textsf{\bfseries 4,5}}, and
  Christian Pfleiderer\textsuperscript{\textsf{\bfseries 1,6,7}}}
  
\mail{e-mail
  \textsf{andreas.bauer@ph.tum.de}}

\institute{%
  \textsuperscript{1}\,Physik-Department, Technische Universit\"at M\"unchen, D-85748 Garching, Germany\\
  \textsuperscript{2}\,Leibniz Institute for Solid State and Materials Research IFW, D-01171 Dresden, Germany\\
  \textsuperscript{3}\,Heinz Maier-Leibnitz Zentrum (MLZ), D-85748 Garching, Germany\\
  \textsuperscript{4}\,Institut für Kristallographie, RWTH Aachen, D-52056 Aachen, Germany\\
  \textsuperscript{5}\,J\"ulich Centre for Neutron Science (JCNS) at Heinz Maier-Leibnitz Zentrum (MLZ), D-85748 Garching, Germany\\
  \textsuperscript{6}\,Munich Center for Quantum Science and Technology (MCQST), Technische Universit\"at M\"unchen, D-85748 Garching, Germany\\
  \textsuperscript{7}\,Zentrum f\"ur QuantumEngineering (ZQE), Technische Universit\"at M\"unchen, D-85748 Garching, Germany}


\abstract{\bf%
  The availability of large high-quality single crystals is an important prerequisite for many studies in solid-state research. The optical floating-zone technique is an elegant method to grow such crystals, offering potential to prepare samples that may be hardly accessible with other techniques. As elaborated in this report, examples include single crystals with intentional compositional gradients, deliberate off-stoichiometry, or complex metallurgy. For the cubic chiral magnets Mn$_{1-x}$Fe$_{x}$Si and Fe$_{1-x}$Co$_{x}$Si, we prepared single crystals in which the composition was varied during growth from $x = 0 - 0.15$ and from $x = 0.1 - 0.3$, respectively. Such samples allowed us to efficiently study the evolution of the magnetic properties as a function of composition, as demonstrated by means of neutron scattering. For the archetypical chiral magnet MnSi and the itinerant antiferromagnet CrB$_{2}$, we grew single crystals with varying initial manganese (0.99 to 1.04) and boron (1.95 to 2.1) content. Measurements of the low-temperature properties addressed the correlation between magnetic transition temperature and sample quality. Furthermore, we prepared single crystals of the diborides ErB$_{2}$, MnB$_{2}$, and VB$_{2}$. In addition to high vapor pressures, these materials suffer from peritectic formation, potential decomposition, and high melting temperature, respectively.}

\maketitle   

\section{Introduction}

The optical floating-zone technique is an elegant method of crystal growth that may be applied to a wide range of materials. In this crucible-free technique, a small molten zone is created by means of optical heating and slowly traversed through feed rods of the material to be grown. As one of the key advantages, besides being crucible-free, this method permits precise control on the stoichiometry of the feed material and the starting composition of the molten zone. The latter enables flux growth approaches which, in conjecture with the relatively small volume of the molten zone, allow to address metallurgically complex phase diagrams and materials with fairly high vapor pressure. In addition, due to the direct optical access to the molten zone, the process parameters may be optimized already during growth.

We have set up a preparation chain that is focused on growth of large high-quality single crystals of intermetallic compounds under ultra-high vacuum compatible conditions. Details on the technical aspects can be found in Refs.~\cite{2011_Neubauer_RevSciInstrum,2016_Bauer_RevSciInstrum,2016_Bauer_RevSciInstruma}. Core parts are three image furnaces that allow us to address a wide range of materials. The furnaces are a four-mirror image furnace from Crystal Systems Corporation that was completely refurbished to be all-metal sealed~\cite{2011_Neubauer_RevSciInstrum}, a four-mirror image furnace from Crystal Systems Corporation with inclinable mirrors that is optimized for large-diameter samples, and a high-pressure high-temperature image furnace from Scientific Instruments Dresden.

In any case, mechanically stable poly-crystalline feed rods with appropriate stoichiometry are required for crystal growth by means of the optical floating-zone technique. For the preparation of such rods, various furnaces were developed in order to meet the specific requirements of the materials to be grown. An argon glove box permits us to handle air-sensitive elements exclusively in an inert atmosphere. Using bespoke load-locks, the starting elements may be transferred into an arc-melting furnace or an induction-heated cold boat furnace. In these systems, the starting material may be fused together. Feed rods are prepared either directly in one of those furnaces or by means of an induction-heated rod casting furnace that is based on an Hukin type cold crucible~\cite{2016_Bauer_RevSciInstruma}. In addition, two annealing furnaces as well as a solid-state electrotransport setup are available for post-growth treatment.

The furnaces are all-metal sealed, bakeable, and may be pumped to ultra-high vacuum using combinations of scroll pumps, turbomolecular pumps, and/or ion getter pumps. As most materials of interest involve elements with high vapor pressure, the vacuum typically just represents the precondition for the application of a high-purity inert atmosphere in form of 6N argon additionally purified by point-of-use gas purifiers.

Following this brief overview on our preparation chain, we showcase studies on two rather different classes of intermetallic compounds. First, single crystals of the cubic chiral magnets Mn$_{1-x}$Fe$_{x}$Si and Fe$_{1-x}$Co$_{x}$Si were grown with an intentional compositional gradient along the growth direction. Using small-angle neutron scattering, these crystals allow us to track the evolution of the magnetic properties as a function of composition over a wide parameter range on a single single-crystalline specimen. Moreover, we briefly recapitulate a recent study using positron spectroscopy on the archetypical chiral magnet MnSi that correlated the initial manganese content with the type and concentration of crystalline point defects and the magnetic properties~\cite{2016_Reiner_SciRep}. 

Second, single crystals of various hexagonal diborides were prepared. In CrB$_{2}$, the influence of the initial boron content on the antiferromagnetic phase transition was studied. For ErB$_{2}$, which forms through a peritectic reaction~\cite{1996_Liao_JPhaseEquilib}, the first large single crystal was grown by means of a traveling-solvent floating-zone approach. Measurements of the low-temperature properties revealed pronounced magnetic anisotropy. For MnB$_{2}$, which in fact is only a metastable compound at room temperature~\cite{1986_Liao_BullAlloyPhaseDiagr}, also the first large single crystal was grown. At low temperatures, a rearrangement of the magnetic structure accompanied by relatively weak anisotropy was observed. For VB$_{2}$, which was used as a nonmagnetic reference, complications due to very the high melting temperature were overcome.

Although belonging to the same class of materials, different approaches for the preparation of poly-crystalline feed rods were required for each diboride. While for CrB$_{2}$ the constituent elements were alloyed in the arc-melting furnace before rods were cast in the rod casting furnace, for ErB$_{2}$ the feed rods were produced exclusively in the arc-melting furnace. For MnB$_{2}$, constituent elements in powder form were sintered using hot tungsten crucibles in the cold boat furnace before rods were cast in a rod casting furnace. For VB$_{2}$, the sintered rods were directly used as feed rods. As the isotope $^{10}$B exhibits a high absorption cross-section for thermal neutrons, all diborides were grown from $^{11}$B enriched boron powder in order to enable efficient neutron scattering studies.

\section{Chiral magnets}

The cubic chiral magnet MnSi has been investigated intensively for more than half a century. During this time, it served as model system for long-wavelength helical structures~\cite{1976_Ishikawa_SolidStateCommun,1976_Motoya_SolidStateCommun}, weak itinerant ferromagnetism~\cite{1979_Moriya_JMagnMagnMater,1985_Lonzarich_JPhysC}, and the possible breakdown of Fermi liquid theory under pressure~\cite{2001_Pfleiderer_Nature,2013_Ritz_Nature}. During the last decade, the skyrmion lattice state, a hexagonal lattice of topologically nontrivial spin whirls emerging in finite magnetic field, has attracted considerable scientific interest~\cite{2009_Muhlbauer_Science,2010_Yu_Nature}.

MnSi and related B20 compounds, such as FeGe or Fe$_{1-x}$Co$_{x}$Si, crystallize in space group $P2_{1}3$ that lacks inversion symmetry. The magnetic properties may be described in terms of a hierarchy of energy scales that comprise in decreasing strength (i)~ferromagnetic exchange interaction, (ii)~Dzyaloshinskii--Moriya interaction as leading-order term in spin--orbit coupling~\cite{1957_Dzialoshinskii_SovPhysJETP,1960_Moriya_PhysRev,1964_Dzyaloshinskii_SovPhysJETP}, and (iii)~higher-order terms in spin--orbit coupling, also referred to as magneto-crystalline anisotropy~\cite{1980_Bak_JPhysC,1981_Plumer_JPhysC,1981_Kataoka_JPhysSocJpn,2017_Bauer_PhysRevB}. This hierarchy is reflected in a rather universal magnetic phase diagram, in which characteristic temperature, field, and length scales may vary considerably for different materials. Furthermore, doped systems, such as Mn$_{1-x}$Fe$_{x}$Si, Fe$_{1-x}$Co$_{x}$Si, or Mn$_{1-x}$Fe$_{x}$Ge, offer the possibility to continuously tune the strength of the interactions over a wide parameter range~\cite{2013_Nagaosa_NatNanotechnol,2016_Bauer_Book}.

The space group $P2_{1}3$ exists in two enantiomers, left-handed and right-handed. Only due to this lack of inversion symmetry, the Dzyaloshinskii--Moriya interactions may arise, thereby inheriting the handedness of the crystalline structure to the magnetic structure~\cite{1983_Shirane_PhysRevB}. The relation between crystalline chirality $\mathit{\Gamma}_{c}$ and magnetic chirality $\gamma_{m}$ is fixed for a given material, also determining the sign of the Dzyaloshinskii--Moriya constant. In MnSi and Mn$_{1-x}$Fe$_{x}$Si one observes $\mathit{\Gamma}_{c}\gamma_{m} = +1$~\cite{1985_Ishida_JPhysSocJpn,1985_Tanaka_JPhysSocJpn,2009_Grigoriev_PhysRevLett,2010_Grigoriev_PhysRevB,2013_Morikawa_PhysRevB} while in Fe$_{1-x}$Co$_{x}$Si it is $\mathit{\Gamma}_{c}\gamma_{m} = -1$~\cite{2009_Grigoriev_PhysRevLett,2011_Dyadkin_PhysRevB,2013_Morikawa_PhysRevB}. Within a doping series, however, the relation also can change as reported for Mn$_{1-x}$Fe$_{x}$Ge. With increasing $x$, the product $\mathit{\Gamma}_{c}\gamma_{m}$ switches from $+1$ in MnGe to $-1$ in FeGe around $x \approx 0.75$, where a vanishingly small Dzyaloshinskii--Moriya constant and ferromagnetic spin alignment are observed~\cite{2013_Grigoriev_PhysRevLett,2013_Shibata_NatNanotechnol}.

Conventional studies as a function of composition can be quite tedious as each data point in a compositional phase diagram usually requires the preparation of a dedicated sample. Specimens prepared with smooth, well-controlled compositional gradients, however, can circumvent such limitations, in particular for congruently melting compounds such as the transition-metal silicides MnSi, FeSi, CoSi, and mixtures thereof~\cite{2010_Okamoto_Book}. When combined with non-destructive experimental techniques suitable to probe large-volume samples with sufficient spatial resolution, such as small-angle neutron scattering or neutron depolarization radiography~\cite{2019_Jorba_JMagnMagnMater}, the evolution of magnetic properties as a function of composition may be tracked efficiently on a single sample.

As we will demonstrate in the following, the optical floating-zone technique is ideally suited for the growth of single crystals with intentional compositional gradients by using feed rods with different compositions. In principle, also more complex composition profiles within the feed rods can be realized. Besides an analysis of the composition by means of energy-dispersive X-ray spectroscopy, both standard and polarized small-angle neutron scattering data are presented. Note that similar concepts also can be exploited in micrometer-sized samples studied by means of Lorentz transmission electron microscopy~\cite{2011_Yu_NatMater,2013_Shibata_NatNanotechnol}.

\subsection{Gradient crystals}

\begin{figure}
	\includegraphics*[width=\linewidth]{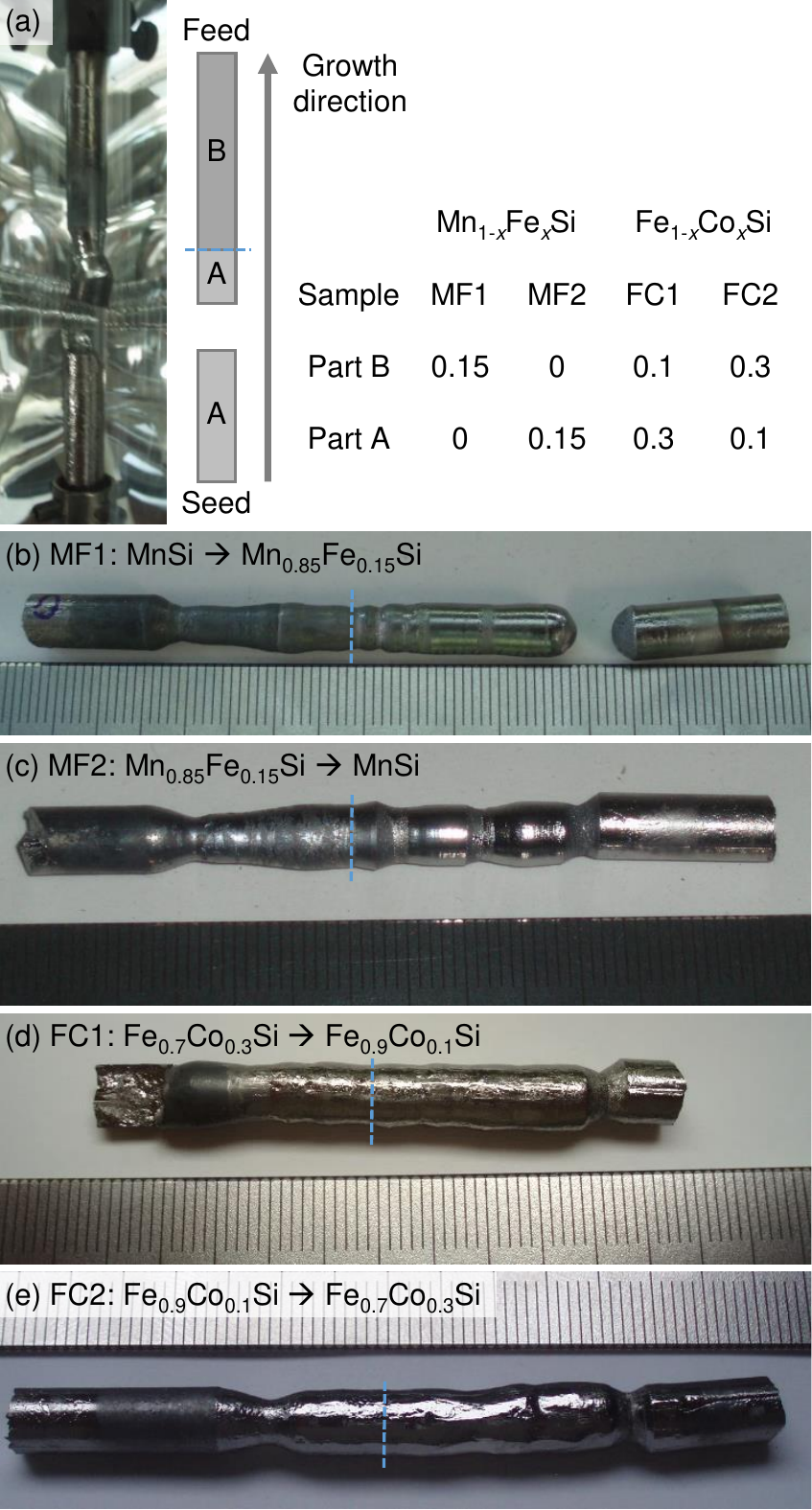}
	\caption{Single-crystal growth of cubic chiral magnets with compositional gradients. (a)~Poly-crystalline seed rod with composition A (bottom) and feed rod with composition B (top) as mounted in the image furnace. Prior to the growth, a small piece of composition A was welded to the lower part of the feed rod using the image furnace. The four samples grown are summarized in the table, where the corresponding values of $x$ are presented. \mbox{(b)--(e)}~Float-zoned ingots. Growth direction was from left to right. The blue dashed line indicates the initial joining point of compositions A and B in the feed rod. The scales are in millimeter.}
	\label{figure01}
\end{figure}

For the growth of single crystals of cubic chiral magnets with compositional gradients, first poly-crystalline rods of the different end compositions were prepared from high-purity starting elements (pre-cast 4N manganese~\cite{2016_Bauer_RevSciInstruma}, 4N iron flakes, 3N5 cobalt pellets, and 6N silicon pieces) using the induction-heated rod casting furnace. Rods with a diameter of 6~mm were cast for Mn$_{1-x}$Fe$_{x}$Si with $x = 0$ and $x = 0.15$ as well as Fe$_{1-x}$Co$_{x}$Si with $x = 0.1$ and $x = 0.3$. In total, four gradient crystals were grown as summarized in Fig.~\ref{figure01}. Samples denoted `MF' and `FC' are composed of Mn$_{1-x}$Fe$_{x}$Si and Fe$_{1-x}$Co$_{x}$Si, respectively. Samples with names including `1' were started from the composition with higher magnetic ordering temperature, while samples with names including `2' were started from the composition with lower magnetic ordering temperature. 

Prior to each floating-zone process, as illustrated in Fig.~\ref{figure01}(a), a small piece of material with the composition of the seed rod, A, was welded to the bottom of the feed rod with composition B. This way, grain selection at the start of the crystal growth process can take place in pristine environment, preventing any irregularities that may arise due to steep compositional gradients. Using the all-metal sealed four-mirror image furnace under a static high-purity argon atmosphere of ${\sim}2$~bar, all crystals were float-zoned at a rate of 5~mm/h while seed and feed rod were counter-rotating at 23~rpm and 10~rpm, respectively.

The resulting ingots are shown in Figs.\ref{figure01}(b) to \ref{figure01}(e), namely sample MF1 going from MnSi to Mn$_{0.85}$Fe$_{0.15}$Si, sample MF2 going from Mn$_{0.85}$Fe$_{0.15}$Si to MnSi, sample FC1 going from Fe$_{0.7}$Co$_{0.3}$Si to Fe$_{0.9}$Co$_{0.1}$Si, and sample FC1 going from Fe$_{0.9}$Co$_{0.1}$Si to Fe$_{0.7}$Co$_{0.3}$Si. Along the growth direction, from left to right in the pictures, the remaining poly-crystalline seed rod with composition A is followed by the start of the growth with the regime of grain selection that was further promoted by a slight necking, with exception of sample FC1. The initial joining point of the starting compositions within the single-crystalline section is marked by a dashed blue line. At the right-hand side, there are the quenched final zone, with exception of sample MF1, and the poly-crystalline feed rod with composition B. 

Single-crystallinity, in particular across the compositional gradient, was confirmed by means of X-ray Laue diffraction as well as neutron diffraction at the single-crystal diffractometer RESI at the Heinz Maier-Leibniz Zentrum~(MLZ)~\cite{2015_Pedersen_JLarge-ScaleResFacil}. Note that due to the use of poly-crystalline feed rods, the growth directions and hence the axes of the single-crystal cylinders were not aligned with any major crystallographic axes.

\begin{figure}
	\includegraphics*[width=\linewidth]{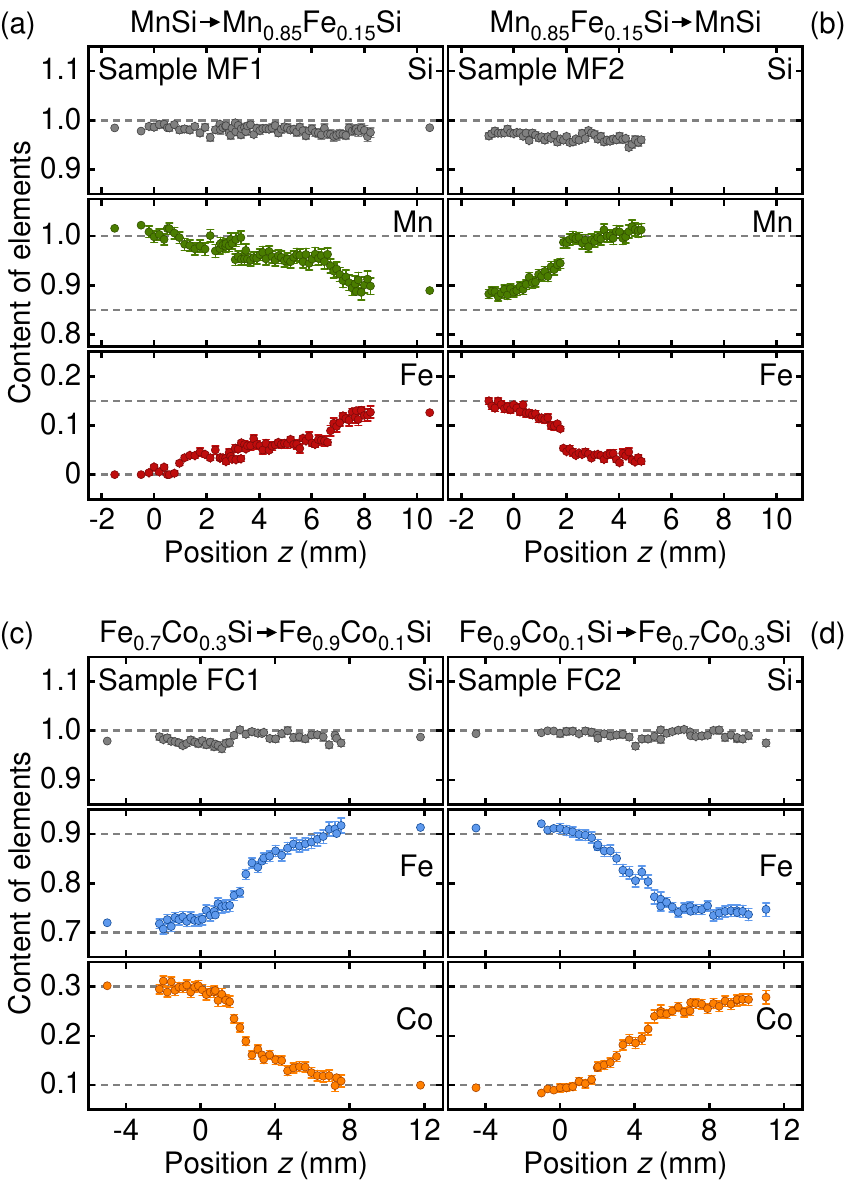}
	\caption{Composition of the float-zoned samples along the growth direction as inferred from energy-dispersive X-ray spectroscopy. The position of the initial joining point of the compositions in the feed rod was defined as $z = 0$. (a)~Sample MF1 from MnSi to Mn$_{0.85}$Fe$_{0.15}$Si. (b)~Sample MF2 from Mn$_{0.85}$Fe$_{0.15}$Si to MnSi. (c)~Sample FC1 from Fe$_{0.7}$Co$_{0.3}$Si to Fe$_{0.9}$Co$_{0.1}$Si. (d)~Sample FC2 from Fe$_{0.9}$Co$_{0.1}$Si to Fe$_{0.7}$Co$_{0.3}$Si.}
	\label{figure02}
\end{figure}

The composition of the single crystals was studied using energy-dispersive X-ray spectroscopy on a Zeiss EVO MA25 scanning electron microscope. For these measurements, one side of each float-zoned ingot was mechanically polished along the growth direction $z$, where $z = 0$ refers to the initial joining point of the starting compositions. On this polished surface, area scans across the entire grown length were combined with line scans in the vicinity of the initial joining point in order to obtain sufficient spatial resolution.

Figure~\ref{figure02} summarizes the evolution of the sample composition along the growth direction as inferred from the X-ray spectroscopy. For all four samples studied, the silicon content (gray symbols) remains essentially constant and close to the expected value of 1. Sample MF1 starts from MnSi, which means at a manganese content of 1 (green symbols) and a vanishing iron content (red symbols). Upon reaching the initial joining point $z = 0$, the iron content starts to increase while the manganese content decreases accordingly. Within about 7~mm, both iron and manganese contents monotonically approach the values expected for the final composition of the feed rod of Mn$_{0.85}$Fe$_{0.15}$Si. Note that the transition from the initial to the final composition essentially takes place within a distance that corresponds to the height of the molten zone during growth. Similarly, in sample MF2 the manganese content increases from 0.85 to 1 within several millimeter, while the iron content decreases from 0.15 to 0. Here, a rather steep change, accounting for about one third of the total change of composition, occurs around $z = 2~\mathrm{mm}$. As no peculiarities were observed during the growth process, the origin of this jump in composition remains unknown.
   
The situation in the Fe$_{1-x}$Co$_{x}$Si samples is comparable. In sample FC1 the iron content (blue symbols) increases from 0.7 to 0.9 within a few millimeter, while the cobalt content (orange symbols) decreases from 0.3 to 0.1. Likewise, in sample FC2 the iron content decreases while the cobalt content increases. Taken together, the results of energy-dispersive X-ray spectroscopy indicate that all four single crystals exhibit a well-controlled monotonic evolution of their composition. Consequently, when studying the magnetic properties of these samples, a direct correlation between the position $z$ along the growth direction and the respective composition is available.

\subsection{Neutron scattering}

\begin{figure}
	\includegraphics*[width=\linewidth]{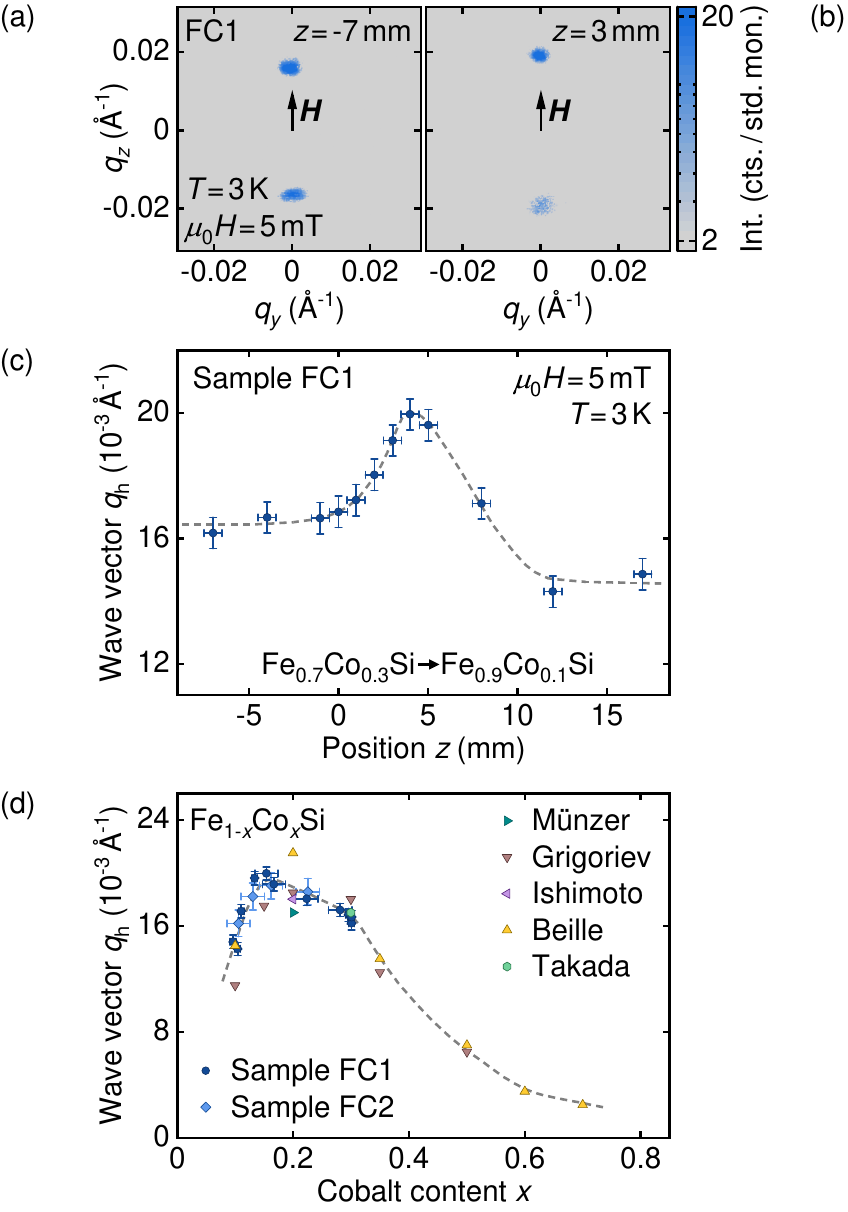}
	\caption{Small-angle neutron scattering on the Fe$_{1-x}$Co$_{x}$Si sample FC1. \mbox{(a),(b)}~Scattering patterns at low temperature recorded in a small vertical magnetic field after high-field cooling for different positions $z$ along the vertically aligned growth direction. (c)~Helical wave vector $q_{\mathrm{h}}$ as a function of the position $z$. (d)~Helical wave vector $q_{\mathrm{h}}$ as a function of the cobalt content $x$ as determined by combining the results of small-angle neutron scattering and energy-dispersive X-ray spectroscopy. For comparison, data are shown from reports by M\"{u}nzer \textit{et al.}~\cite{2010_Munzer_PhysRevB}, Grigoriev \textit{et al.}~\cite{2009_Grigoriev_PhysRevLett}, Ishimoto \textit{et al.}~\cite{1995_Ishimoto_PhysicaB}, Beille \textit{et al.}~\cite{1983_Beille_SolidStateCommun}, and Takada \textit{et al.}~\cite{2009_Takeda_JPhysSocJpn}.}
	\label{figure03}
\end{figure}

The magnetic properties of the single crystals grown with compositional gradients were studied by means of small-angle neutron scattering on the multi-purpose instrument MIRA at MLZ~\cite{2015_Georgii_JLarge-ScaleResFacil,2018_Georgii_NuclInstrumMethodsPhysResA}. The samples were mounted with their growth direction aligned vertically in cryostats providing base temperatures of 450~mK for Mn$_{1-x}$Fe$_{x}$Si and 3~K for Fe$_{1-x}$Co$_{x}$Si. Magnetic fields along the vertical $z$ axis were generated by a pair of water-cooled Helmholtz coils, i.e., the magnetic field was not aligned with a major crystallographic axes. Rocking scans were possible around the horizontal $y$ axis that was perpendicular to both the magnetic field and the neutron beam. Using an incident neutron wavelength of 9.7~\AA\ as well as source and sample apertures of $5\times1~\mathrm{mm}^{2}$ and $5\times0.5~\mathrm{mm}^{2}$, respectively, the height of the neutron beam was limited to below 1~mm. This spatial resolution of about 1~mm allowed us to study the evolution of the magnetic order when successively scanning along the growth direction $z$ of the samples.

Prior to the measurements, the samples FC1 and FC2 were high-field cooled in a vertical magnetic field of 180~mT, before the field was reduced to 5~mT. In a doped chiral magnet such as Fe$_{1-x}$Co$_{x}$Si, such a field and temperature history yields a single-domain state oriented along the field direction~\cite{1995_Ishimoto_PhysicaB,2007_Grigoriev_PhysRevB,2010_Munzer_PhysRevB,2016_Bauer_PhysRevB,2020_Kindervater_PhysRevB}. The samples MF1 and MF2 were measured in magnetic fields of 110~mT and 180~mT, respectively~\cite{2010_Bauer_PhysRevB}. Consistent with the literature, no magnetic peaks were observed in Mn$_{1-x}$Fe$_{x}$Si for iron concentrations $x > 0.1$~\cite{2011_Grigoriev_PhysRevB,2020_Kindervater_PhysRevB}.

Typical scattering patterns as recorded at two different positions $z$ on sample FC1 are shown in Figs.~\ref{figure03}(a) and \ref{figure03}(b). One intensity maximum is observed each along the positive and the negative $z$ direction, corresponding to the field direction. The distance between each maximum and the origin defines the helical wave vector $q_{\mathrm{h}}$, as characteristic for a single-domain helimagnetic state with a real-space modulation length of $2\pi/q_{\mathrm{h}}$. 

When measuring at different positions $z$ along growth direction of the sample, the size of the helical wave vector changes with the composition, as summarized in Fig.~\ref{figure03}(c). Starting from a value of about 0.016~\AA$^{-1}$ for Fe$_{0.7}$Co$_{0.3}$Si, a maximum emerges within the regime of the compositional gradient, before the wave vector again saturates at about 0.015~\AA$^{-1}$ for Fe$_{0.9}$Co$_{0.1}$Si. This behavior is in excellent agreement with previous reports\cite{2010_Munzer_PhysRevB,2009_Grigoriev_PhysRevLett,1995_Ishimoto_PhysicaB,1983_Beille_SolidStateCommun,2009_Takeda_JPhysSocJpn}, which becomes particularly clear when combing the results of small-angle neutron scattering and energy-dispersive X-ray spectroscopy, as shown in Fig.~\ref{figure03}(d). 

The helical wave vector in Fe$_{1-x}$Co$_{x}$Si exhibits a distinct maximum. The present study allows us to locate this maximum rather precisely at $x = 0.15$, demonstrating the potential of samples with intentional compositional gradient especially for the determination of compositional phase diagrams. Compared to conventional studies, a large number of data points may be inferred from a single sample. In turn, concentrations of interest for further studies may be selected accurately, for instance when expensive starting materials are involved or the growth of large high-quality single crystals is required for inelastic neutron scattering experiments. 

\begin{figure}
	\includegraphics*[width=\linewidth]{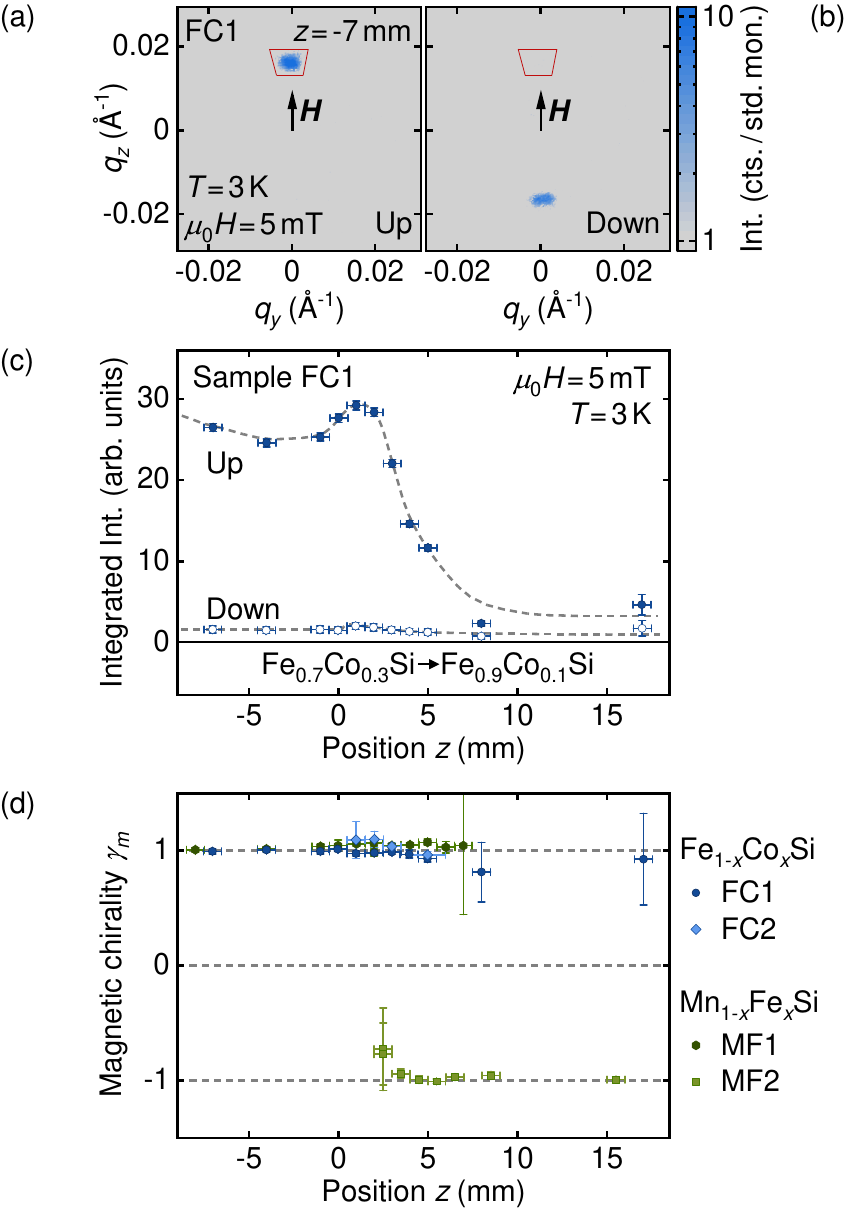}
	\caption{Polarized small-angle neutron scattering on the Fe$_{1-x}$Co$_{x}$Si sample FC1. \mbox{(a),(b)}~Scattering patterns at low temperature under a small magnetic field after high-field cooling. The incident neutron beam was polarized either parallel to the positive $z$ direction (Up, left) or parallel to the negative $z$ direction (Down, right). (c)~Integrated intensity of the maximum at positive $q_{z}$ values, cf.\ red box in panels (a) and (b), as a function of the position $z$ along the growth direction. (d)~Magnetic chirality $\gamma_{m}$ along the growth direction $z$ for the four samples studied.}
	\label{figure04}
\end{figure}

In addition to standard small-angle neutron scattering, the four samples available were also studied by means of polarized small-angle neutron scattering. For these experiments, a polarizing supermirror was used to polarize the incoming neutron beam along the positive vertical $z$ direction, in the following denoted as `up'. Using a spin flipper located between the source and the sample aperture, the polarization of the incoming neutron beam could be flipped, in the following denoted as `down'. As shown in Figs.~\ref{figure04}(a) and \ref{figure04}(b), for sample FC1 only the scattering maximum at positive $q_{z}$ is observed in the up configuration, while only the maximum at negative $q_{z}$ is observed in the down configuration. This selectivity is a direct consequence of the handedness of the helix in the cubic chiral magnets. 

Akin to the analysis in Ref.~\cite{2009_Grigoriev_PhysRevLett}, the magnetic chirality $\gamma_{m}$ may be determined as 
\begin{equation}
	\label{equation1}
	\gamma_{m} = \frac{ I_{\mathrm{up}} - I_{\mathrm{down}}} {I_{\mathrm{up}} + I_{\mathrm{down}}}.
\end{equation}
Here, the values $I_{\mathrm{up}}$ and $I_{\mathrm{down}}$ are inferred from integrating the intensity around the helical satellite at (0,$+q_{h}$) in the area shown as red box in Figs.~\ref{figure04}(a) and \ref{figure04}(b). $I_{\mathrm{up}}$ is the intensity when the incoming neutron beam is polarized parallel to the positive $z$ axis, while $I_{\mathrm{down}}$ is intensity when the polarization is parallel to the negative $z$ axis. When also accounting for the background intensity $I_{\mathrm{BG}}$ and the flipping ratio $R$ of the spin flipper, Eq.~\eqref{equation1} expands to 
\begin{equation}
	\label{equation2}
	\gamma_{m} = \frac{I_{\mathrm{up}} - I_{\mathrm{down}} + I_{\mathrm{up}}/R}{I_{\mathrm{up}} + I_{\mathrm{down}} - 2I_{\mathrm{BG}} - I_{\mathrm{up}}/R}.
\end{equation}

For each sample, the magnetic chirality stays unchanged, when analyzing the polarized scattering data at all sample positions yielding sufficient intensity, as shown in Fig.~\ref{figure04}(c). This observation is consistent with the literature, which expects the product of crystalline chirality $\mathit{\Gamma}_{c}$ and magnetic chirality $\gamma_{m}$ to be $+1$ for helimagnetic Mn$_{1-x}$Fe$_{x}$Si and $-1$ for helimagnetic Fe$_{1-x}$Co$_{x}$Si. In fact, the magnetic chirality is directly linked to the crystalline chirality which in turn is determined when one of the grains with random handedness becomes dominant during grain selection. Note that a particular crystalline handedness can be induced by picking a corresponding seed crystal~\cite{2011_Dyadkin_PhysRevB}. Taken together, a magnetic chirality $\gamma_{m} = +1$ for FC1, FC2, and MF1 as well as $\gamma_{m} = -1$ for MF2 corresponds to a crystalline chirality $\mathit{\Gamma}_{c} = +1$ for MF1 as well as $\mathit{\Gamma}_{c} = -1$ for FC1, FC2, and MF2.

\subsection{Positron spectroscopy}

\begin{figure}
	\includegraphics*[width=\linewidth]{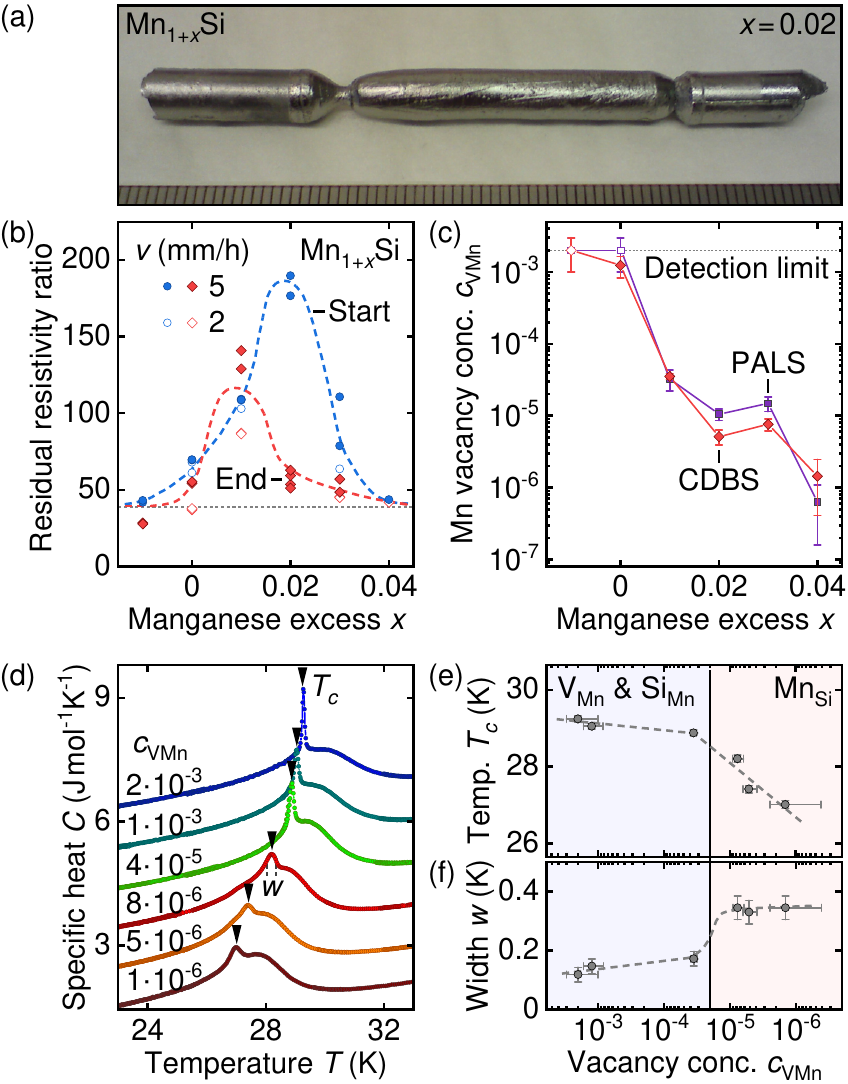}
	\caption{Single-crystal growth of the chiral magnet Mn$_{1+x}$Si with varying initial manganese excess $x$. (a)~Typical optically float-zoned ingot. Growth direction was from left to right. The scale is in millimeter. (b)~Residual resistivity ration, RRR, as a function of the initial manganese excess. Crystals grown with different growth rates, $v$, are compared. Samples prepared from the beginning and end of the single-crystal ingots are denoted `start' and `end', respectively. (c)~Concentration of manganese vacancies, $c_{\mathrm{VMn}}$, as inferred from coincident Doppler broadening spectroscopy~(CDBS) and positron annihilation lifetime spectroscopy~(PALS) as a function of the initial manganese excess. (d)~Specific heat as a function of temperature for samples with different concentration of manganese vacancies. \mbox{(e),(f)}~Transition temperature, $T_{c}$, and the width of the transition, $w$, as a function of the concentration of manganese vacancies.}
	\label{figure05}
\end{figure}

The precise control of the stoichiometry of the poly-crystalline feed rods also allowed us to study the influence of different types of point defects on the magnetic properties of the archetypical chiral magnet MnSi by means of positron spectroscopy. This study was reported in detail in Ref.~\cite{2016_Reiner_SciRep} and is recapitulated briefly in the following. Starting from pre-cast 4N manganese rods and 6N silicon pieces, poly-crystalline rods of Mn$_{1+x}$Si with an initial manganese excess $x = -0.01, 0, 0.01, 0.02, 0.03, 0.04$ were prepared in the induction-heated rod casting furnace. Single-crystal growth of MnSi, congruently melting at $1270^{\circ}\mathrm{C}$~\cite{1991_Okamoto_JPhaseEquilib}, was carried out in the all-metal sealed four-mirror image furnace under a static high-purity argon atmosphere of 1.5~bar at growth rates of 2~mm/h and 5~mm/h, respectively, while seed and feed rod were counter-rotating at 6~rpm. Prior to each melting process, the furnaces were pumped down to ${\sim}1\times10^{-7}$~mbar. As indicated by means of X-ray Laue diffraction, all growth attempts yielded single crystals with a diameter of 6~mm and a length of 10~mm to 30~mm, where a typical float-zoned ingot is depicted in Fig.~\ref{figure05}(a). No preferred direction of growth was observed. 

The residual resistivity ratio RRR, corresponding to the ratio of the electrical resistivity at room temperature to the resistivity at low temperature, was determined for samples cut from the beginning and the end of each single crystal, denoted `start' and `end', respectively. As shown in Fig.~\ref{figure05}(b), the highest values were observed for a small initial excess of manganese. In metallic specimens, high RRR values are typically associated with small total concentrations of defects, consistent with the compensation of evaporation of manganese during growth by means of a small initial excess. In the parameter range studied, the growth rate had no influence on the RRR.

In order to study the point defects in further detail, two positron spectroscopy techniques were used, notably coincident Doppler broadening spectroscopy and positron annihilation lifetime spectroscopy~\cite{1996_Asoka-Kumar_PhysRevLett,2000_Coleman_Book,2012_Hugenschmidt_NewJPhys}. Both technique permitted us to identify the type and density of vacancies in the single crystals and yielded consistent results. As shown in Fig.~\ref{figure05}(c), with increasing initial manganese excess, the concentration of manganese vacancies V$_{\textrm{Mn}}$ decreases from at least $2\cdot10^{-3}$, corresponding to the upper detection limit due to saturation trapping, to $1\cdot10^{-6}$. Note that no silicon vacancies V$_{\textrm{Si}}$ were detected. These experimental results were compared with ab-initio calculations of effective formation energies of point defects, indicating that for small manganese contents silicon atoms on manganese sites Si$_{\textrm{Mn}}$ and manganese vacancies V$_{\textrm{Mn}}$ dominate, while for large manganese contents manganese atoms on silicon sites Mn$_{\textrm{Si}}$ prevail. The smallest absolute density of point defects, associated with the highest RRR, is expected for a manganese to silicon ratio of exactly 1:1.

As shown in terms of the specific heat in Fig.~\ref{figure05}(d), the magnetic properties are qualitatively robust but nevertheless are influenced by certain types of point defects rather than by the RRR, as reported previously~\cite{2001_Pfleiderer_JMagnMagnMater,2011_Stishov_Phys-Usp,2012_Bauer_PhysRevB,2016_Ou-Yang_JPhysCondensMatter}. With decreasing concentration of manganese vacancies $c_{\mathrm{VMn}}$, the transition temperature $T_{c}$ shifts to smaller values and the anomaly at the transition distinctly broadens, as summarized in Figs.~\ref{figure05}(e) and \ref{figure05}(f). While this evolution may seem counter-intuitive at first glance, the ab-initio calculations indicate that a decreasing concentration of manganese vacancies is accompanied by an increasing concentration of manganese atoms occupying silicon sites, Mn$_{\textrm{Si}}$. 

The combination of these findings suggests that manganese atoms on silicon sites are particularly efficient in modifying the electronic structure and the magnetic interactions in MnSi when compared to other types of point defects. Hence, while the RRR may reflect the absolute density of defects, precise knowledge on the type of defects may be key when characterizing magnetic or electronic properties in complex materials such as MnSi. In this context, single crystals prepared with deliberate off-stoichiometry represent the foundation for studies exploring the interplay of defects and phenomena such as the current-induced depinning of skyrmions~\cite{2010_Jonietz_Science,2012_Yu_NatCommun,2013_Iwasaki_NatCommun,2013_Lin_PhysRevB,2015_Muller_PhysRevB} or the dynamics of their topological unwinding~\cite{2013_Milde_Science,2017_Wild_SciAdv}.

\section{Diborides}

The class of C32 diborides comprises a large number of transition-metal and rare-earth intermetallic compounds that crystallize in the hexagonal space group $P6/mmm$. The rather simple crystal structure consists of closest-packed metal layers stacked with honeycomb boron layers along the hexagonal $c$ direction. The diborides were long known for their high mechanical, thermal, and chemical stability, but huge scientific interest in this class of compounds was generated in 2001 when conventional two-band superconductivity with a record-high transition temperature of 39.5~K was discovered in MgB$_{2}$~\cite{2001_Nagamatsu_Nature,2003_Canfield_PhysToday}.

In addition to superconductivity, a wide range of electronic and magnetic properties are realized in diborides, depending on the metal atom. Consequently, phenomena such as quantum phase transitions or electronic topological transitions may be addressed by means of compositional tuning within the same simple crystal structure. Despite this potential, in particular studies of single-crystalline material remain scarce for many diborides. In this context, challenges in the preparation process may play a crucial role, including peritectic or peritectoid reactions and high melting temperatures that intimately connect to high vapor pressures and evaporation losses. In the following, we report on the single-crystal growth of CrB$_{2}$, ErB$_{2}$, MnB$_{2}$, and VB$_{2}$, establishing the optical floating-zone technique as a potent tool for the growth of refractory compounds with potentially complex metallurgy.

CrB$_{2}$ has been studied comparably well, thereby establishing this material as an itinerant antiferromagnet par excellence with a N\'{e}el temperature of 88~K~\cite{1969_Barnes_PhysLettA,2009_Grechnev_JAlloyCompd,2014_Bauer_PhysRevB}. Hallmark properties include, good metallic conductivity~\cite{1976_Guy_JPhysChemSolids,1976_Tanaka_JLessCommonMet}, small ordered moments of $0.5~\mu_{\mathrm{B}}/$Cr~\cite{1977_Funahashi_SolidStateCommun} contrasted by large fluctuating moments of $2~\mu_{\mathrm{B}}/$Cr~\cite{1966_Cadeville_JPhysChemSolids,1976_Tanaka_JLessCommonMet,2005_Balakrishnan_JCrystGrowth}, and an exceptional stability with respect to magnetic fields, all arising in the presence of moderate electronic correlations and sizable geometric frustration~\cite{1969_Castaing_SolidStateCommun,2013_Brasse_PhysRevB,2014_Bauer_PhysRevB}. An incommensurate cycloidal magnetic structure was inferred from neutron scattering~\cite{1977_Funahashi_SolidStateCommun,2009_Kaya_PhysicaB} that, however, is not fully consistent with nuclear magnetic resonance data~\cite{1978_Kitaoka_SolidStateCommun,1980_Kitaoka_JPhysSocJpn,2007_Michioka_JMagnMagnMater}. Recently, a comprehensive neutron scattering study, to be reported elsewhere~\cite{2019_Regnat_PhD}, potentially connected the complex magnetic structure to boron vacancies. In order to investigate this aspect further, however, samples with different concentrations of such vacancies are required.

For ErB$_{2}$, only very few studies are available. Early on, magnetic order below 16~K was inferred from susceptibility data measured on arc-melted poly-crystals and attributed to ferromagnetism due to a positive Curie--Weiss temperature~\cite{1977_Buschow_Book}. More recently, a kink in the electrical resistivity at 14~K characteristic of the onset of magnetic order~\cite{2015_Kargin_Book} and an increase of the lattice constant $c$ below about 100~K~\cite{2010_Novikov_PhysSolidState} were reported. For these studies, poly-crystalline samples prepared by means of multi-step processes involving high-pressure synthesis were used that contained several percent of impurity phases~\cite{2009_Matovnikov_InorgMater,2010_Novikov_PhysSolidState}. Reports on single-crystalline material so far were limited to the growth of platelets of $1\times1\times0.01~\mathrm{mm}^{3}$ from an excess rare-earth flux~\cite{1972_Castellano_MaterResBull}. Note that there also no comprehensive studies for related rare-earth diborides, such as DyB$_{2}$, HoB$_{2}$, or TmB$_{2}$.

In MnB$_{2}$, early studies reported the emergence of small ferromagnetic moments of ${\sim}0.2~\mu_{\mathrm{B}}/$Mn at temperatures below about 150~K. In conjecture with large fluctuating moments of more than $2~\mu_{\mathrm{B}}/$Mn, this finding was interpreted as a sign for weak itinerant ferromagnetism~\cite{1966_Andersson_SolidStateCommuna,1966_Cadeville_JPhysChemSolids,1969_Barnes_PhysLettA}. Shortly after, nuclear magnetic resonance and neutron powder diffraction revealed that MnB$_{2}$ in fact is a large-moment antiferromagnet. Its N\'{e}el temperature lies well above room temperature and the magnetic moments point mostly perpendicular to the $c$ axis~\cite{1969_Kasaya_JPhysSocJpna,1970_Kasaya_JPhysSocJpn,1972_Legrand_SolidStateCommun}. Later on, band structure calculations indicated the local-moment character of this antiferromagnetism and suggested that the ferromagnetic moment at low temperatures arises from a small spin canting~\cite{2000_Khmelevskyi_SolidStateCommun,2012_Khmelevskyi_JPhysCondensMatter}. So far, single crystals were limited to platelets of $0.5\times0.5\times0.05~\mathrm{mm}^{3}$ grown by sublimation from sintered material in a sealed silica tube~\cite{1970_Kasaya_JPhysSocJpn}. For the paramagnetic VB$_{2}$~\cite{1972_Castaing_JPhysChemSolids}, large single crystals were grown using an induction-heated floating-zone approach~\cite{1981_Nakano_JCrystGrowth}.

\subsection{Chromium diboride}

\begin{figure}
	\includegraphics*[width=\linewidth]{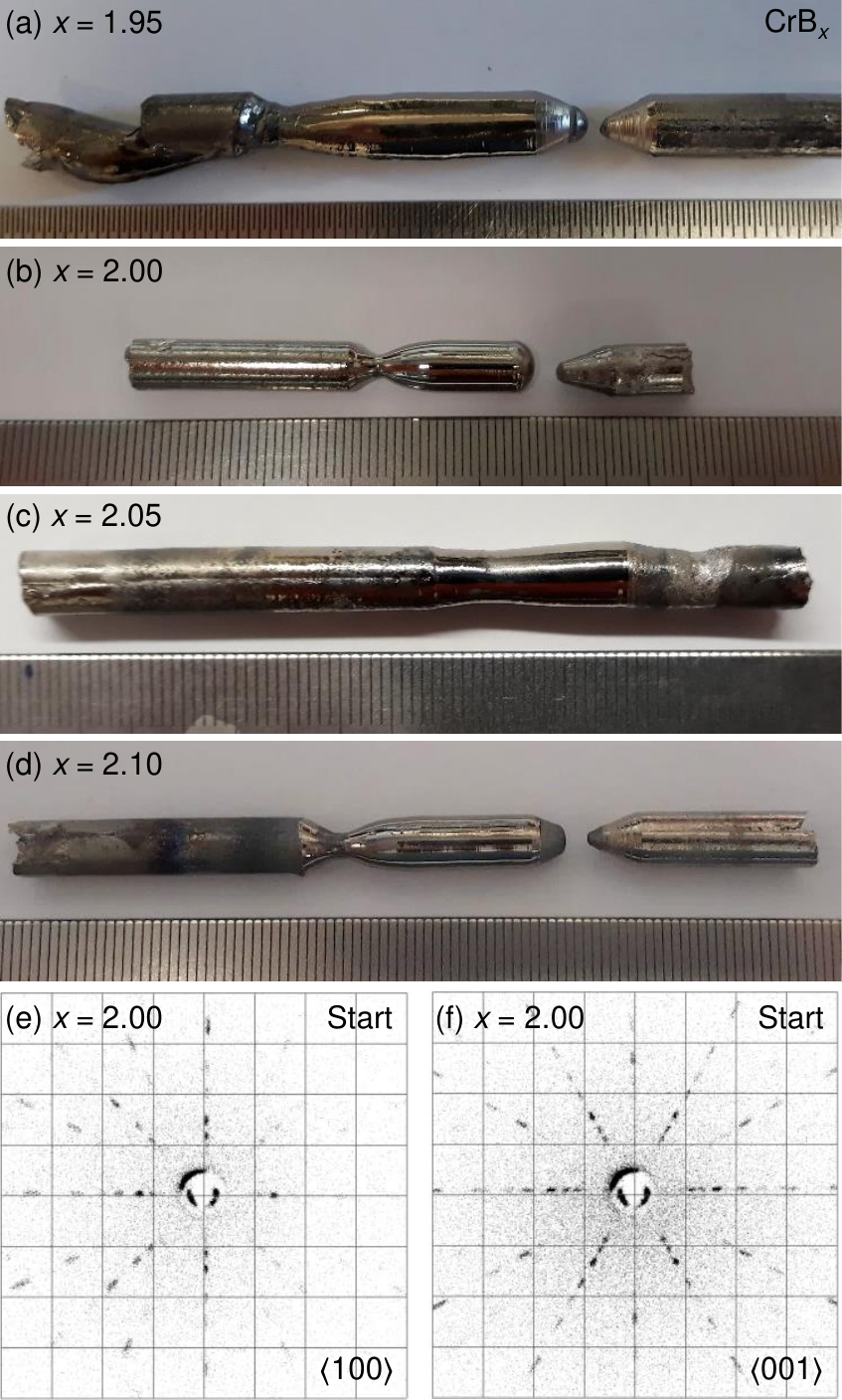}
	\caption{Single-crystal growth of the hexagonal diboride CrB$_{x}$ with varying initial boron content $x$. \mbox{(a)--(d)}~Optically float-zoned ingots. Growth direction was from left to right. The scales are in millimeter. \mbox{(e),(f)}~X-ray Laue diffraction patterns along the twofold-symmetric $\langle100\rangle$ axis and the sixfold-symmetric $\langle001\rangle$ axis.}
	\label{figure06}
\end{figure}

The transition-metal diboride CrB$_{2}$ melts congruently~\cite{1986_Liao_BullAlloyPhaseDiagra}, but both chromium and boron exhibit considerable vapor pressure at the high melting temperature of ${\sim}2200^{\circ}\mathrm{C}$, which means that evaporation losses during growth can be hardly avoided. In order to account for this circumstance and investigate the putative influence of boron vacancies on the magnetic properties of CrB$_{2}$, a series of single crystals was grown with initial boron contents ranging from 1.95 to 2.1, akin to the study on the chiral magnet MnSi presented above~\cite{2016_Reiner_SciRep}.

The poly-crystalline feed rods of CrB$_{x}$ were prepared in a two-step process. First, appropriate amounts of 5N chromium granules and 5N boron coarse powder (98\% $^{11}$B enriched) were fused together in the arc-melting furnace. Each batch comprised about 3~g of material and was homogenized by repeated cycles of melting, cooling, and turning over. Second, several of the resulting pills were loaded into the induction-heated rod casting furnace, in which a poly-crystalline rod with a diameter of 6~mm was prepared. Both a shorter seed and a longer feed rod were cast for each composition CrB$_{x}$ studied, namely $x = 1.95$, $2.00$, $2.05$, and $2.10$. Prior to each melting process, the furnaces were pumped down to ${\sim}1\times10^{-6}$~mbar and subsequently flooded with about 1~bar of 6N argon additionally purified by a point-of-use gas purifier.

The single crystal growth was carried out in the high-pressure high-temperature floating-zone furnace, where an argon atmosphere of ${\sim}15$~bar was applied, flowing at a rate of 0.1~l/min. The crystals were float-zoned at a rate of 5~mm/h while seed and feed rod were counter-rotating. Despite distinct evaporation losses, stable growth conditions were obtained during all growth attempts. Figure~\ref{figure06} shows the four float-zoned ingots of CrB$_{x}$, where the growth direction was from left to right. After grain selection, typically promoted by a necking, all growth attempts yielded single-crystalline cylinders of 10~mm to 20~mm height, as confirmed by means of X-ray Laue diffraction. Typical Laue patterns recorded on the sample with $x = 2.00$ are shown in Fig.~\ref{figure06}(e) for the twofold-symmetric $a$ axis and in Fig.~\ref{figure06}(f) for the sixfold-symmetric $c$ axis.

From each float-zoned ingot, using a wire saw, a set of samples was cut from both beginning and end of the single crystal, denoted `start' and `end', respectively. This set consisted of (i)~a piece with a mass of ${\sim}30$~mg from which powder was prepared for X-ray diffractometry, (ii)~a platelet of $2\times1\times0.2~\mathrm{mm}^{3}$ with orientations along $\langle001\rangle\times\langle210\rangle\times\langle100\rangle$ for measurements of the electrical resistivity, and (iii)~a cuboid of $2.5\times2.2\times1~\mathrm{mm}^{3}$ with orientations along $\langle001\rangle\times\langle100\rangle\times\langle210\rangle$ for measurements of the specific heat.

\begin{figure}
	\includegraphics*[width=\linewidth]{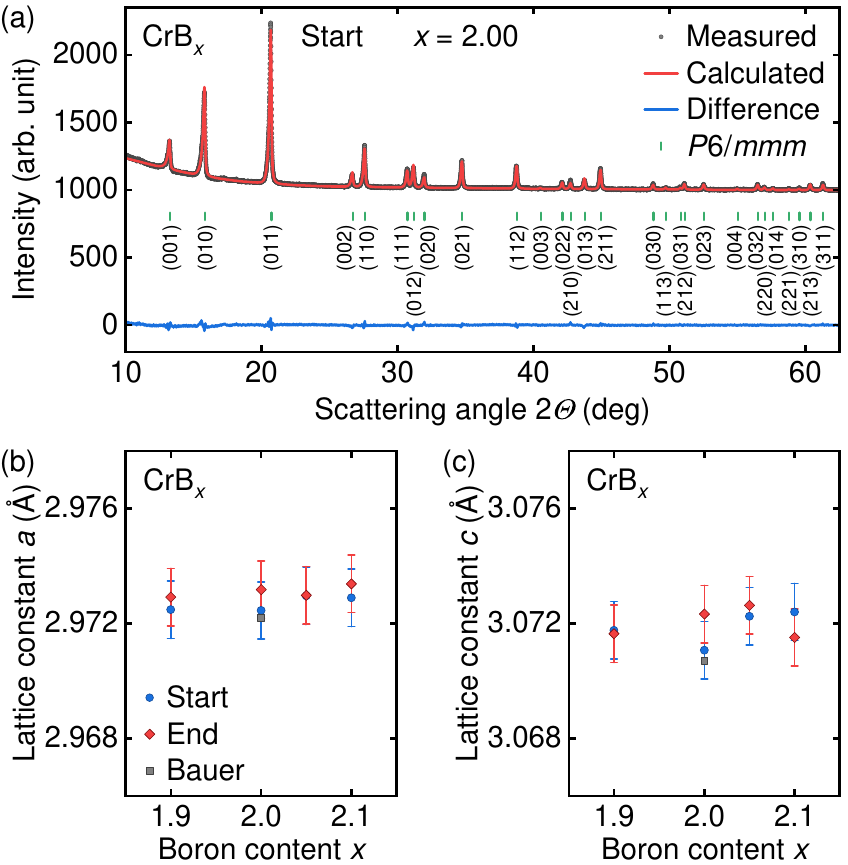}
	\caption{X-ray diffraction and lattice constants of CrB$_{x}$. (a)~X-ray powder diffractogram of float-zoned material for $x = 2.00$. Measured data are in excellent agreement with a Rietveld refinement based on the hexagonal $P6/mmm$ structure. \mbox{(b),(c)}~Lattice constants $a$ and $c$ inferred from Rietveld refinements as a function of the initial boron content $x$. Samples from start and end of the single-crystal ingots were investigated. For comparison, the values from Bauer \textit{et al.}~\cite{2014_Bauer_PhysRevB} are shown.}
	\label{figure07}
\end{figure}

The structural properties of CrB$_{x}$ were studied using a Huber G670 X-ray powder diffractometer in Guinier geometry with a Mo source. As an example for a typical diffractogram, data for the specimen from the start of the crystal with $x = 2.00$ are shown in Fig.~\ref{figure07}(a). A Rietveld refinement based on the hexagonal space group $P6/mmm$ is in excellent agreement with the data. Lattice constants $a = 2.973$~\AA\ and $c = 3.071$~\AA\ are inferred, in good agreement with previous reports~\cite{1987_Nishihara_JPhysSocJpn,2001_Vajeeston_PhysRevB,2014_Bauer_PhysRevB}.

For all initial boron contents $x$ studied, the X-ray diffractograms are consistent with phase-pure CrB$_{2}$ crystallizing in space group $P6/mmm$. As shown in Figs.~\ref{figure07}(b) and \ref{figure07}(c), within the estimated error bars of $1\times10^{-3}$~\AA\ the lattice constants $a$ and $c$ remain unchanged as a function of $x$.

\begin{figure}
	\includegraphics*[width=\linewidth]{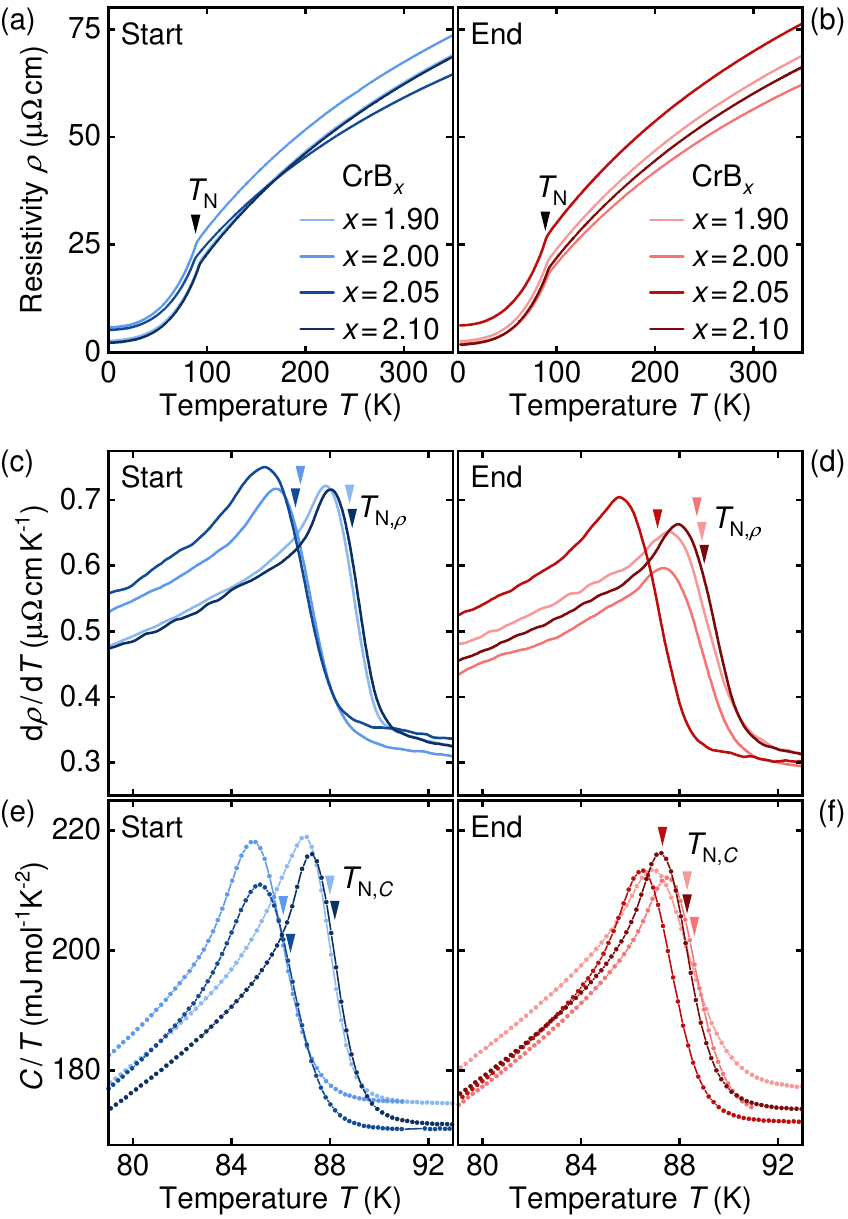}
	\caption{Electrical resistivity and specific heat of CrB$_{x}$. Samples cut from start and end of the single-crystal ingots were studied for different initial boron content $x$. \mbox{(a),(b)}~Electrical resistivity $\rho$ as a function of temperature. The onset of antiferromagnetic order at the N\'{e}el temperature $T_{\mathrm{N}}$ translates into a pronounced change of slope. \mbox{(c),(d)}~Derivative of the resistivity, $\mathrm{d}\rho/\mathrm{d}T$, around the antiferromagnetic phase transition. \mbox{(e),(f)}~Specific heat divided by temperature, $C/T$, around the antiferromagnetic phase transition.}
	\label{figure08}
\end{figure}

The low-temperature properties of CrB$_{x}$ were studied using a Quantum Design physical property measurement system. For the electrical resistivity, $\rho$, samples were contacted in a four-terminal configuration using a wire bonder. The measurements were carried out by means of a custom setup based on lock-in amplifiers (Stanford Research 830) and a dedicated current source (Keithley 6221). The specific heat $C$ was measured using a standard small-heat pulse method, where heat pulses had a size of 2\% of the current temperature.

\begin{table}[b]
	\caption{Summary of key parameter inferred for CrB$_{x}$. The residual resistivity ratio RRR corresponds to the ratio of the resistivity at room temperature, $\rho_{300\,\mathrm{K}}$, to the resistivity at low temperature, $\rho_{4\,\mathrm{K}}$. The magnetic transition temperatures inferred from measurements of the resistivity and the specific are denoted $T_{\mathrm{N},\rho}$ and $T_{\mathrm{N},C}$, respectively.}
	\begin{tabular}[htbp]{@{}lllllll@{}}
		\hline
		Sample  &       & $\rho_{300\,\mathrm{K}}$   & $\rho_{4\,\mathrm{K}}$     & RRR  & $T_{\mathrm{N},\rho}$ & $T_{\mathrm{N},C}$ \\
		$x$     &       & $\mathrm{\mu\Omega\,cm}$   & $\mathrm{\mu\Omega\,cm}$   &      & K                     & K                  \\
		\hline
		1.90    & Start & 62.7                       & 2.7                        & 23.2 & 88.8                  & 88.0               \\
		1.90    & End   & 62.6                       & 2.6                        & 24.1 & 88.9                  & 88.3               \\
		2.00    & Start & 67.4                       & 5.9                        & 11.4 & 86.8                  & 86.1               \\
		2.00    & End   & 56.5                       & 2.0                        & 28.3 & 88.7                  & 88.6               \\
		2.05    & Start & 59.1                       & 5.3                        & 11.2 & 86.6                  & 86.4               \\
		2.05    & End   & 70.0                       & 6.3                        & 11.1 & 87.1                  & 87.3               \\
		2.10    & Start & 62.2                       & 2.3                        & 27.0 & 88.9                  & 88.2               \\
		2.10    & End   & 60.1                       & 1.8                        & 33.4 & 89.0                  & 88.3               \\
		\hline
	\end{tabular}
	\label{table1}
\end{table}

In Figs.~\ref{figure08}(a) and \ref{figure08}(b) the electrical resistivity is shown for all samples studied from room temperatures down to 2~K. Samples prepared from the start and the end of the single-crystal ingots are shown in the left and the right column, respectively. All samples exhibit typical metallic behavior with a distinct change of slope at the onset of antiferromagnetic order at the N\'{e}el temperature $T_{\mathrm{N}}$, consistent with previous reports~\cite{1976_Tanaka_JLessCommonMet,2014_Bauer_PhysRevB}. The absolute values of the resistivity vary considerably with no clear trend as a function of $x$. As summarized in Table~\ref{table1}, the same holds true for the residual resistivity ratio RRR, for which geometric effects cancel out. Low residual resistivities and therefore high RRR values, which in metallic samples are typically associated with high sample purity, are observed for the smallest and highest values of $x$ as well as for the end of the crystal grown with $x = 2.00$.

The influence of the initial boron content $x$ on the antiferromagnetic phase transition is illustrated by means of the derivative of the resistivity around the N\'{e}el temperature, shown in Figs.~\ref{figure08}(c) and \ref{figure08}(d). Defining the N\'{e}el temperature in spirit of an entropy-conserving construction, as described in Ref.~\cite{2014_Bauer_PhysRevB}, transition temperatures are inferred that again follow no clear trend as a function of $x$. The properties inferred at the start and the end of the single-crystal ingot are comparable. The specific heat divided by temperature, $C/T$, shown in Figs.~\ref{figure08}(e) and \ref{figure08}(f), is highly reminiscent of the derivative of the resistivity, suggesting that both quantities are governed by the same energy scales. The N\'{e}el temperatures inferred from the specific heat are consistent with the values obtained from the resistivity data, in particular when considering that different samples were measured. 

All N\'{e}el temperatures are summarized in Table~\ref{table1}. No clear dependence on the initial boron content $x$ is observed, but instead a correlation between RRR and N\'{e}el temperatures is established. Samples with high RRR exhibit high N\'{e}el temperatures, around 88.5~K, while samples with low RRR exhibit low N\'{e}el temperatures, between 86~K and 87.5~K. This finding indicates that the crystalline quality possesses a direct impact on the itinerant antiferromagnetism in CrB$_{2}$, also providing a potential explanation for slightly different transition temperatures reported throughout the literature. The crystalline quality in turn is influenced by various parameters during growth and not just a simple function of the initial boron content.

\subsection{Erbium diboride}

\begin{figure}
	\includegraphics*[width=\linewidth]{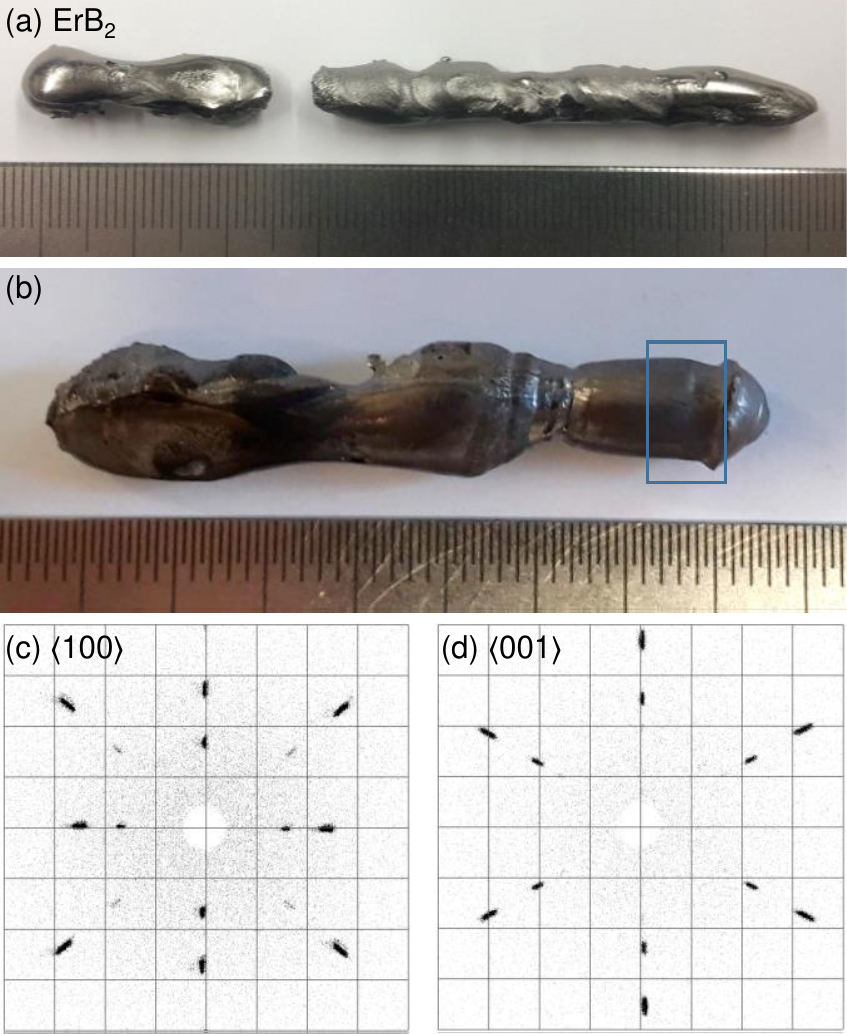}
	\caption{Single-crystal growth of the hexagonal diboride ErB$_{2}$. (a)~Poly-crystalline seed and feed rods as prepared using the arc-melting furnace. (b)~Optically float-zoned ingot. Growth direction was from left to right. As ErB$_{2}$ forms through a weakly peritectic reaction, the self-adjusted flux traveling-solvent floating-zone approach was used. At the end of the process, marked by the blue box, a large single crystal was obtained. The scales are in millimeter. \mbox{(c),(d)}~X-ray Laue diffraction patterns along the twofold-symmetric $\langle100\rangle$ axis and the sixfold-symmetric $\langle001\rangle$ axis.}
	\label{figure09}
\end{figure}

The rare-earth diboride ErB$_{2}$ forms through a weakly peritectic reaction, which means that the composition of liquid end point of the peritectic reaction (${\sim}65$ atomic percent boron) just differs slightly from the composition of the target compound (66.7 atomic percent boron)~\cite{1996_Liao_JPhaseEquilib}. This small difference makes ErB$_{2}$ an ideal candidate for single-crystal growth by means of the self-adjusted flux traveling-solvent floating-zone approach. For this technique, seed and feed rods are prepared with the target stoichiometry ErB$_{2}$. When creating the molten zone from the tips of both rods at about $2250~^{\circ}\mathrm{C}$ and slowly traversing it through the feed rod, the melt precipitates boron-rich material, probably in form of ErB$_{4}$. This way, the composition of the molten zone is shifted towards smaller boron content and its melting temperature decreases until the liquid end point of the peritectic reaction is reached at ${\sim}65$ atomic percent boron and $2185~^{\circ}\mathrm{C}$.

Starting materials were 4N erbium pieces and 5N boron coarse powder (98\% $^{11}$B enriched) that were exclusively handled inside the argon glove box due to the air-sensitivity of rare-earth elements such as erbium. From there, appropriate amounts of the starting materials were loaded into the arc-melting furnace and fused to pills with a mass of about 3~g. Subsequently, several of these pills were welded together in the arc-melting furnace, giving rise to a rod-shaped ingot. Prior to each melting process, the furnace was pumped down to ${\sim}1\times10^{-6}$~mbar and flooded with about 1~bar of 6N argon additionally purified by a point-of-use gas purifier.

Two ingots were prepared, depicted in Fig.~\ref{figure09}(a), that served as poly-crystalline seed and feed rod for a single-crystal growth in the high-pressure high-temperature floating-zone furnace. In order to reduce evaporation losses, an argon atmosphere of ${\sim}18$~bar was applied, flowing at a rate of 0.1~l/min. A relatively fast growth rate of 3~mm/h was applied as the requirements of the self-adjusted flux technique had to be balanced against the sizable vapor pressures of both erbium and boron at temperatures well above $2000~^{\circ}\mathrm{C}$. Seed and feed rod were counter-rotating at 10~rpm. In order to keep the volume of the molten zone constant, no necking was introduced.

The float-zoned ingot is depicted in Fig.~\ref{figure09}(b). During the first ${\sim}10$~mm of growth, the molten zone adjusted and grain selection took place, resulting in large single-crystalline grains in the final ${\sim}10$~mm of the ingot as marked by the blue box. These grains had a size of several millimeter in all dimensions and their good crystalline quality was indicated by X-ray Laue patterns shown in Fig.~\ref{figure09}(c) for the twofold-symmetric $a$ axis and in Fig.~\ref{figure09}(d) for the sixfold-symmetric $c$ axis. Note that preliminary tests by means of X-ray powder diffraction on arc-melted specimens suggested that the peritectically forming diborides DyB$_{2}$ and HoB$_{2}$ are also suitable candidates for the self-adjusted flux traveling-solvent floating-zone approach, while for TmB$_{2}$ evaporation losses may pose a serious challenge~\cite{2010_Okamoto_Book}.

\begin{figure}
	\includegraphics*[width=\linewidth]{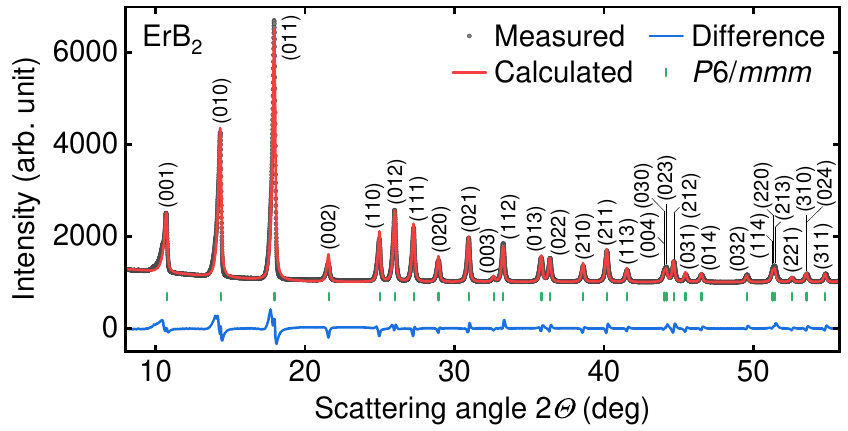}
	\caption{X-ray powder diffractogram of float-zoned material of ErB$_{2}$. Measured data are in excellent agreement with a Rietveld refinement based on the hexagonal $P6/mmm$ structure.}
	\label{figure10}
\end{figure}

An X-ray powder diffractogram of ErB$_{2}$ recorded on a Huber G670 diffractometer is shown in Fig.~\ref{figure10}. The data are in excellent agreement with Rietveld refinement based on the hexagonal space group $P6/mmm$. Lattice constants $a = 3.275$~\AA\ and $c = 3.784$~\AA were inferred, consistent with previous reports~\cite{1977_Buschow_Book,1996_Liao_JPhaseEquilib,2010_Novikov_PhysSolidState}. No indications of impurity phases, such as ErB$_{4}$, are observed that might by associated with the peritectic formation of ErB$_{2}$, indicating that the sample is phase-pure.

\begin{figure}
	\includegraphics*[width=\linewidth]{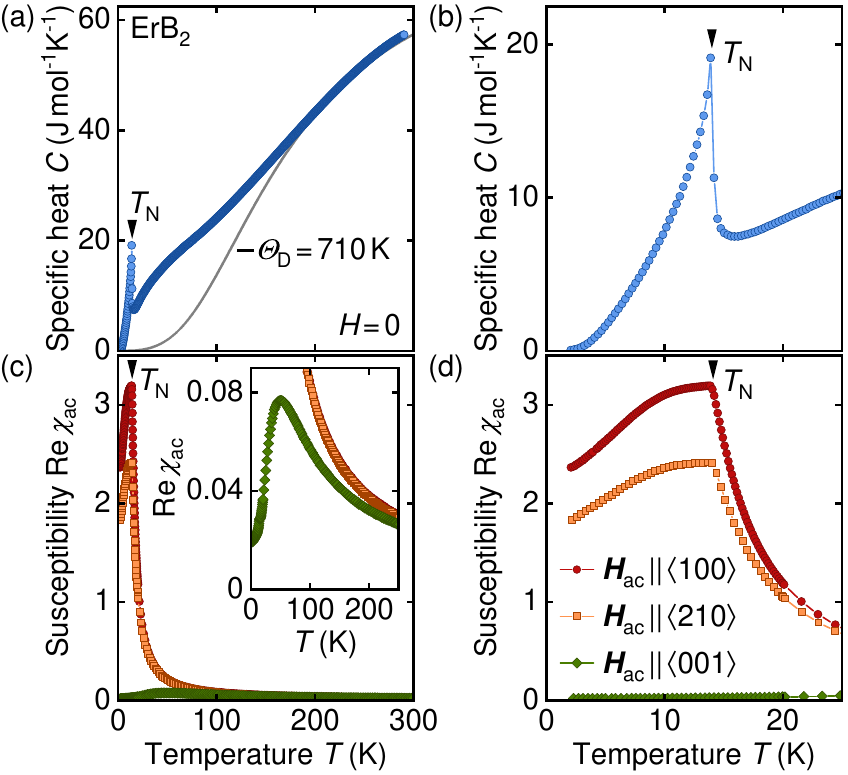}
	\caption{Specific heat and ac susceptibility of ErB$_{2}$. (a)~Specific heat as a function of temperature. A pronounced peak is observed at the onset of antiferromagnetic order at the N\'{e}el temperature $T_{\mathrm{N}}$. As an estimate for the phonon contribution, a Debye model with $\mathit{\Theta}_{\mathrm{D}} = 710~\mathrm{K}$ is shown (solid gray line). (b)~Specific heat around the antiferromagnetic phase transition. (c)~Real part of the ac susceptibility, $\mathrm{Re}\,\chi_{\mathrm{ac}}$, as a function of temperature for excitation fields along the major crystallographic axes. Inset: Enlarged view of the susceptibility for excitation field along the hard $\langle001\rangle$ direction. (d)~Ac susceptibility around the antiferromagnetic phase transition.}
	\label{figure11}
\end{figure}

For measurements of the low-temperature properties of ErB$_{2}$, a cuboid of $2.5\times1.4\times1~\mathrm{mm}^{3}$ with orientations along $\langle001\rangle\times\langle100\rangle\times\langle210\rangle$ was cut from a larger single-crystal ingot using a wire saw. In a Quantum Design physical property measurement system, the specific heat was measured using a quasi-adiabatic large heat pulse technique~\cite{2013_Bauer_PhysRevLett} and the ac susceptibility was measured with an excitation field of 1~mT at an excitation frequency of 1~kHz.

The specific heat, $C$, as a function of temperature is shown in Fig.~\ref{figure11}(a). At high temperatures, the specific heat slowly approaches the Dulong--Petit limit of $74.8~\mathrm{J\,mol}^{-1}\mathrm{K}^{-1}$. For comparison, the specific of the Debye model calculated for a Debye temperature $\mathit{\Theta}_{\mathrm{D}} = 710~$K is shown as gray solid line. Similar to other transition-metal and rare-earth diborides~\cite{2007_Novikov_JThermAnalCalorim,2009_Mori_PhysRevB,2014_Bauer_PhysRevB}, this simple model can serve only as a first rough estimate of the phonon contribution. At low temperatures, a distinct anomaly in the specific heat is associated with the onset of magnetic order at $T_{\mathrm{N}} = 14$~K, consistent with the literature~\cite{2015_Kargin_Book}. As shown in Fig.~\ref{figure11}(b), the low-temperature anomaly exhibits a lambda-like shape, characteristic of a second-order phase transition. Note that in addition the anomaly at $T_{\mathrm{N}}$ a broad shoulder around 50~K is observed, suggesting the emergence of magnetic correlations or spin fluctuations well above $T_{\mathrm{N}}$.

The real part of the ac susceptibility, $\mathrm{Re}\,\chi_{\mathrm{ac}}$, as a function of temperature is shown in Fig.~\ref{figure11}(c) for excitation fields parallel to major crystallographic axes. The absolute value of the susceptibility is high for excitation along $\langle100\rangle$ and $\langle210\rangle$ and rather low for excitation along $\langle001\rangle$, indicating magnetism with strong easy-plane anisotropy. At high temperatures the susceptibility exhibits a Curie-Weiss-like temperature dependence, from which fluctuating moments may be inferred that are consistent with the free-ion value of Er$^{3+}$ of $9.58~\mu_{\mathrm{B}}/$Er. A pronounced kink is observed at the onset of magnetic order at $T_{\mathrm{N}}$ for excitation withing the easy basal plane, as highlighted in Fig.~\ref{figure11}(d). The susceptibility for excitation along the hard $c$ axis lacks a pronounced signature at $T_{\mathrm{N}}$, but displays a broad maximum around 50~K, consistent with the specific heat. Further details on the magnetic properties of ErB$_{2}$ will be reported elsewhere~\cite{2021_Benka_PhD}.

\subsection{Manganese diboride}

\begin{figure}
	\includegraphics*[width=\linewidth]{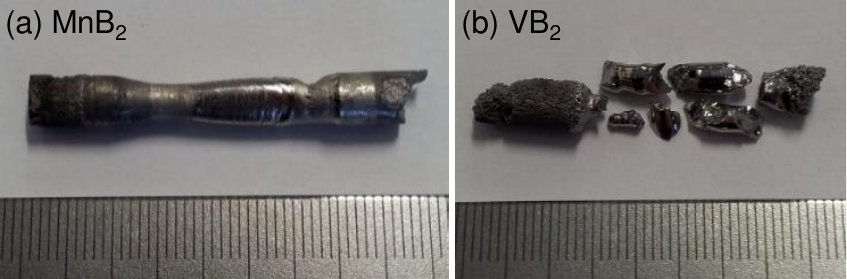}
	\caption{Single-crystal growth of the hexagonal diborides MnB$_{2}$ and VB$_{2}$. (a)~Optically float-zoned ingot of MnB$_{2}$. (b)~Optically float-zoned ingot of VB$_{2}$. Due to the very high melting temperature, no stable zone was accomplished. Still strongly textured material could be obtained. The scales are in millimeter.}
	\label{figure12}
\end{figure}

The transition-metal diboride MnB$_{2}$ melts congruently at $1827~^{\circ}\mathrm{C}$ but in equilibrium is expected to peritectoidly decompose into Mn$_{3}$B$_{4}$ and MnB$_{4}$ around ${\sim}1100~^{\circ}\mathrm{C}$~\cite{1986_Liao_BullAlloyPhaseDiagr}. During a growth by means of the optical floating-zone technique, grain selection can take place quickly, resulting in the formation of a large-volume single crystal at elevated temperatures, which in turn efficiently inhibits the decomposition during cool down. This way, the hexagonal high-temperature phase MnB$_{2}$ can be preserved down to room temperature and below.

For the growth, first, high-purity manganese powder was prepared from 4N manganese rods pre-cast in the induction-heated rod casting furnace~\cite{2016_Bauer_RevSciInstruma}. Stoichiometric amounts of this powder and 4N5 boron powder (99\% $^{11}$B enriched) were thoroughly mixed and sintered at ${\sim}1300~^{\circ}\mathrm{C}$ using a hot tungsten crucible in the induction-heated cold boat furnace. At the IFW Dresden, the resulting material was cast into poly-crystalline rods using an induction-heated rod casting furnace. Two of these rods were mounted to the so-called Smart Floating-Zone furnace, which essentially corresponds to a prototype of the high-pressure furnace from Scientific Instruments Dresden. In order to reduce evaporation losses, the growth was carried out in an argon atmosphere of 80~bar at a rate of 6~mm/h. The float-zoned ingot is shown in Fig.~\ref{figure12}(a). After a few millimeter of grain selection, a single crystal across the entire ingot is observed. In contrast to the other diborides, the growth direction corresponded to the crystallographic $a$ axis and the float-zoned crystal is flattened along the $c$ axis. Further note that the $c$ axis rotates by ${\sim}0.4~\mathrm{deg/mm}$ along the growth direction, as inferred from X-ray Laue diffraction. Across a sample of a size in the millimeter range, as it is used in investigation of low-temperature properties, such rotation is typically negligible.

For the growth of the congruently melting VB$_{2}$~\cite{1987_Spear_JPhaseEquilib}, seed and feed rods were prepared from a stoichiometric mixture of 2N8 vanadium powder and 4N5 boron powder (99\% $^{11}$B enriched) that was sintered at ${\sim}1800~^{\circ}\mathrm{C}$ in a hot tungsten crucible in the induction-heated cold boat furnace. During a single-crystal growth attempt in the Smart Floating-Zone furnace, however, no stable growth conditions were obtained due to the very high melting temperature of ${\sim}2750~^{\circ}\mathrm{C}$. Nevertheless, the resulting ingot, shown in Fig.~\ref{figure12}(b), yielded several samples of millimeter size that were strongly textured, containing large grains with a small orientational mismatch of a few degrees. Such samples can already be sufficient when using VB$_{2}$ as a nonmagnetic, metallic reference compound for specific-heat measurements on magnetic diboride. For further details on the growth process, we refer to Ref.~\cite{2014_Bauer_PhysRevB}.

\begin{figure}
	\includegraphics*[width=\linewidth]{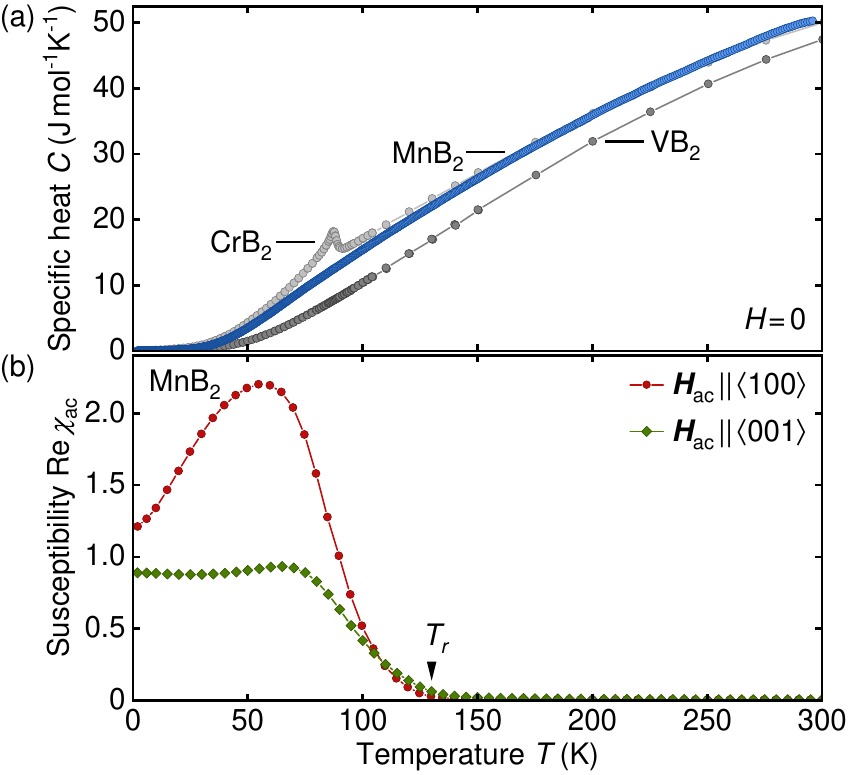}
	\caption{Specific heat and ac susceptibility of MnB$_{2}$. (a)~Specific heat as a function of temperature. Antiferromagnetic order sets in well above room temperature and no anomalies are observed between 2~K and 300~K. For comparison, specific heat data are shown also for the low-temperature antiferromagnet CrB$_{2}$ and the paramagnet VB$_{2}$ as published in Ref.~\cite{2014_Bauer_PhysRevB}. (b)~Ac susceptibility as a function of temperature for excitation fields along the hexagonal $\langle100\rangle$ and $\langle001\rangle$ axes. A distinct increase is observed below a temperature $T_{r} \approx 130~\mathrm{K}$ characteristic of a rearrangement of the magnetic structure.}
	\label{figure13}
\end{figure}

For measurements of the low-temperature properties of MnB$_{2}$, a cuboid of $2.8\times2.7\times1.7~\mathrm{mm}^{3}$ with orientations along $\langle210\rangle\times\langle100\rangle\times\langle001\rangle$ was cut from the large single crystal using a wire saw. In a Quantum Design physical property measurement system, the specific heat was measured using a quasi-adiabatic large heat pulse technique~\cite{2013_Bauer_PhysRevLett} and the ac susceptibility was measured with an excitation field of 1~mT at an excitation frequency of 911~Hz.

The specific heat as a function of temperature, shown in Fig.~\ref{figure13}(a), lacks any signatures characteristic of a phase transition below room temperature. Similar to other diborides, the nonmagnetic contribution can not be described satisfactorily by a simple Debye model. Instead, the specific heat of the low-temperature antiferromagnet CrB$_{2}$ and the paramagnet VB$_{2}$ are shown for comparison, cf.\ Ref.~\cite{2014_Bauer_PhysRevB}. Interestingly, with exception of the pronounced anomaly around its antiferromagnetic phase transition, CrB$_{2}$ essentially tracks the behavior of MnB$_{2}$.

\begin{table}[b]
	\caption{Summary of key parameters for the growth of diborides as used in this study. Besides the initial compositions prior to floating zone process and whether the material forms congruently melting (cong.) or through a peritectic reaction (peri.), the formation temperature $T_{m}$, the pressure of the inert argon atmosphere $p$, and the growth rate $v$ are stated. Feed and seed rods were counter-rotating at rates of about 10~rpm. For VB$_{2}$, no stable molten zone was obtain. In equilibrium MnB$_{2}$ is expected to pertectoidly decompose around $1100~^{\circ}\mathrm{C}$. The nature of single-crystalline (sc) material obtain is briefly summarized in the final column. See text for details.}
	\begin{tabular}[htbp]{@{}llllll@{}}
		\hline
		             &        & $T_{m}$              & $p$ & $v$  & result           \\
		             &        & $^{\circ}\mathrm{C}$ & bar & mm/h &                  \\
		\hline
		CrB$_{1.90}$ & cong.  & 2200                 & 15  & 5    & large sc         \\
		CrB$_{2.00}$ & cong.  & 2200                 & 15  & 5    & large sc         \\
		CrB$_{2.05}$ & cong.  & 2200                 & 15  & 5    & large sc         \\
		CrB$_{2.10}$ & cong.  & 2200                 & 15  & 5    & large sc         \\
		ErB$_{2}$    & peri.  & 2185                 & 18  & 3    & sc grains        \\
		MnB$_{2}$    & cong.  & 1827                 & 80  & 6    & sc with twist    \\
		VB$_{2}$     & cong.  & 2750                 & -   & -    & textured grains  \\
		\hline
	\end{tabular}
	\label{table2}
\end{table}

In contrast to the specific heat, the temperature dependence of the ac susceptibility, shown in Fig.~\ref{figure13}(b), exhibits a strong increase at temperatures below $T_{r} \approx 130~\mathrm{K}$ for excitation fields applied both within the basal plane and perpendicular to it. The absolute values for in-plane excitation are larger by a factor of roughly two, indicating weak easy-plane anisotropy, consistent with field-dependent measurements reported in Ref.~\cite{1970_Kasaya_JPhysSocJpn}. The lack of noticeable signatures in the specific heat, however, implies that the rearrangement of the magnetic structure takes place as a smooth crossover rather than a first-order or second-order phase transition. Further details on the magnetic properties of MnB$_{2}$, including comprehensive neutron diffraction data, will be reported elsewhere~\cite{2019_Regnat_PhD}. Finally, for convenience, Table~\ref{table2} summarizes key parameters used for the growth of diborides in this study.

\section{Conclusion}
In summary, the potential of our preparation chain based on the optical floating-zone technique was demonstrated in terms of the growth of several intermetallic compounds. Precise control over the composition of the feed material allows for studies that utilize single crystals with intentional compositional gradients, showcased for the chiral magnets Mn$_{1-x}$Fe$_{x}$Si and Fe$_{1-x}$Co$_{x}$Si, or deliberate off-stoichiometry, showcased for the chiral magnet MnSi and the itinerant antiferromagnet CrB$_{2}$. Moreover, the preparation chain permits us to handle materials that suffer from complex metallurgical phase diagrams or otherwise challenging growth conditions, such as high vapor pressures and high melting temperatures, as demonstrated by the growth of further diborides, namely ErB$_{2}$, MnB$_{2}$, and VB$_{2}$.

\begin{acknowledgement}
We wish to thank Wolfgang Anwand, Peter B\"{o}ni, Georg Brandl, Maik Butterling, Katarzyna Danielewicz, Andreas Erb, Christian Franz, Thomas Gigl, Saskia Gottlieb-Sch\"{o}nmeyer, Klaudia Hradil, Christoph Hugenschmidt, Florian Jonietz, Petra Kudejova, Michael Leitner, Andreas Mantwill, Susanne Mayr, Sebastian M\"{u}hlbauer, Kirill Nemkovski, Markus Reiner, Julia Repper, Barbara Russ,  Claudia Schweiger, Reinhard Schwikowski, Michael Stanger, Yixi Su, Andreas Wagner, and Marc Wilde for fruitful discussions and assistance with the experiments. This work has been funded by the Deutsche Forschungsgemeinschaft (DFG, German Research Foundation) under TRR80 (From Electronic Correlations to Functionality, Project No.\ 107745057, Project E1), SPP2137 (Skyrmionics, Project No.\ 403191981, Grant PF393/19), and the excellence cluster MCQST under Germany's Excellence Strategy EXC-2111 (Project No.\ 390814868). Financial support by the European Research Council (ERC) through Advanced Grants No.\ 291079 (TOPFIT) and No.\ 788031 (ExQuiSid) is gratefully acknowledged.
\end{acknowledgement}

\bibliographystyle{pss}

\providecommand{\WileyBibTextsc}{}
\let\textsc\WileyBibTextsc
\providecommand{\othercit}{}
\providecommand{\jr}[1]{#1}
\providecommand{\etal}{~et~al.}

\newpage

\section*{Graphical Table of Contents\\}
GTOC image:
\begin{figure}[h]%
\includegraphics[width=4cm,height=4cm]{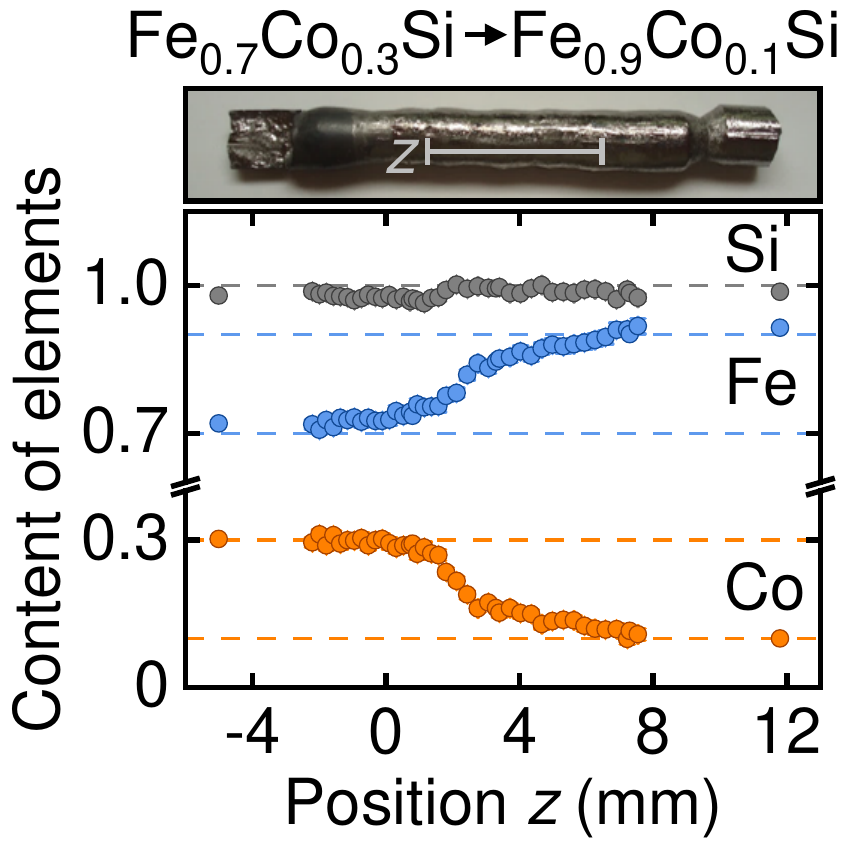}
\caption*{We demonstrated the potential of the optical floating-zone technique for the growth of intermetallic compounds. Notably, we have prepared single crystals of cubic chiral magnets with an intentional compositional gradient along the growth direction (Mn$_{1-x}$Fe$_{x}$Si, Fe$_{1-x}$Co$_{x}$Si), single crystals with deliberate off-stoichiometry (Mn$_{1+x}$Si, $x = -0.01 - 0.04$; CrB$_{x}$, $x = 1.95 - 2.10$), and single crystals of compounds with complex metallurgy or otherwise challenging growth conditions (ErB$_{2}$, MnB$_{2}$, VB$_{2}$).}
\label{GTOC}
\end{figure}

\end{document}